\documentclass[english]{emulateapj}

\usepackage[T1]{fontenc}
\usepackage[latin1]{inputenc}
\usepackage{amssymb}
\usepackage{graphicx}
\usepackage{latexsym}
\usepackage{ae,aecompl}
\usepackage{apjfonts}

\begin{document}

\title{Fast directional correlation on the sphere with steerable filters}

\author{Y. Wiaux\altaffilmark{1}}

\altaffiltext{1}{yves.wiaux@epfl.ch}

\affil{Signal Processing Institute, Ecole Polytechnique Fédérale de Lausanne
(EPFL), CH-1015 Lausanne, Switzerland}

\author{L. Jacques}

\affil{Institut de Physique théorique, Université catholique de Louvain
(UCL), B-1348 Louvain-la-Neuve, Belgium}

\author{P. Vielva}

\affil{Instituto de F\'isica de Cantabria (CSIC-UC), E-39005 Santander,
Spain; and Astrophysics Group, Cavendish Laboratory, University of
Cambridge, CB3 0HE Cambridge, United Kingdom}

\and{}

\author{P. Vandergheynst}

\affil{Signal Processing Institute, Ecole Polytechnique Fédérale de Lausanne
(EPFL), CH-1015 Lausanne, Switzerland}

\begin{abstract}
A fast algorithm is developed for the directional correlation of scalar
band-limited signals and band-limited steerable filters on the sphere.
The asymptotic complexity associated to it through simple quadrature
is of order $\mathcal{O}(L^{5})$, where $2L$ stands for the square-root
of the number of sampling points on the sphere, also setting a band
limit $L$ for the signals and filters considered. The filter steerability
allows to compute the directional correlation uniquely in terms of
direct and inverse scalar spherical harmonics transforms, which drive
the overall asymptotic complexity. The separation of variables technique
for the scalar spherical harmonics transform produces an $\mathcal{O}(L^{3})$
algorithm independently of the pixelization. On equi-angular pixelizations,
a sampling theorem introduced by Driscoll and Healy implies the exactness
of the algorithm. The equi-angular and HEALPix implementations are
compared in terms of memory requirements, computation times, and numerical
stability. The computation times for the scalar transform, and hence
for the directional correlation, of maps of several megapixels on
the sphere ($L\simeq10^{3}$) are reduced from years to tens of seconds
in both implementations on a single standard computer. These generic
results for the scale-space signal processing on the sphere are specifically
developed in the perspective of the wavelet analysis of the cosmic
microwave background (CMB) temperature ($T$) and polarization ($E$
and $B$) maps of the WMAP and Planck experiments. As an illustration,
we consider the computation of the wavelet coefficients of a simulated
temperature map of several megapixels with the second Gaussian derivative
wavelet.
\end{abstract}

\keywords{cosmology: cosmic microwave background --- methods: data analysis
--- methods: numerical}

\section{Introduction}

The cosmic microwave background (CMB) temperature and polarization
anisotropy distribution represents today an exceptional cosmological
laboratory for the study of the geometrical structure, energy content,
and evolution of the universe. The last fifteen years of detection
and analysis, culminating with the all-sky maps of the Wilkinson Microwave
Anisotropy Probe (WMAP) satellite mission, has led to a rather precise
determination of the corresponding cosmological parameters, defining
the concordance cosmological model \cite{CMBpage,CMBspergel1,CMBspergel2}.
The universe is assumed to be globally homogeneous and isotropic (cosmological
principle) and the CMB anisotropies arise from Gaussian quantum fluctuations
defined in a primordial inflationary era. From the signal processing
point of view, the CMB is therefore a unique realization of a Gaussian
and stationary (homogeneous and isotropic) random process on the celestial
sphere and may be completely studied in terms of its temperature and
polarization angular power spectra. The concordance cosmological model
is defined by the values of the cosmological parameters giving the
angular power spectra which best fit the experimental data. However,
among other assumptions, the basic theoretical hypotheses of Gaussianity
and stationarity of the signal must imperatively be thoroughly tested
before the concordance model may be trusted. Many approaches have
been proposed in this direction. In particular, the analysis of local
features in the CMB temperature and polarization anisotropies might
reveal either signatures of fundamental non-Gaussianity or statistical
anisotropy, or foreground emissions. The scale-space analysis of the
CMB, \emph{i.e.} the combined analysis of scale and localization of
features in the signal, is therefore of fundamental interest. In that
perspective, the efficiency of the wavelet signal processing of the
CMB has already been established. First results have also been obtained
from the wavelet analysis of the WMAP all-sky temperature data, notably
in \cite{WNGvielva,WNGmukherjee,WNGmcewen1,WNGcruz}.

A practical approach to wavelets on the sphere was recently introduced
in that context of scale-space analysis of the CMB \cite{SASwiaux1}.
In the proposed formalism, the analysis of a signal is performed through
its wavelet coefficients, resulting from the directional correlation
of the signal with a localized filter to which dilations, rotations,
and translations may be applied. Analogously to the corresponding
formalism on the plane, wavelet filters on the sphere must satisfy
an admissibility condition which guarantees that any signal may be
explicitly reconstructed from its wavelet coefficients. A correspondence
principle was also established, which states that the inverse stereographic
projection of a wavelet on the plane gives a wavelet on the sphere.
This correspondence principle enables to transfer properties, such
as the filter directionality and steerability, from the plane onto
the sphere. Any non-axisymmetric filter is said to be directional.
The good directionality of a filter may be measured in terms of the
peakedness of its auto-correlation function. The steerability of a
filter is that property through which an arbitrary rotation of the
filter around itself may be computed exactly in terms of a finite
number of basis filters \cite{SASwiaux1}.

The present work develops a fast algorithm for the directional correlation
of band-limited signals and band-limited steerable filters on the
sphere, in particular wavelet filters. In the context of the scale-space
CMB analysis, this algorithm reduces typical computation times for
the directional correlation of all-sky maps of several megapixels
from the ongoing WMAP or the forthcoming Planck missions from years
to tens of seconds, thus rendering them easily affordable. The scale-space
wavelet analysis of the CMB temperature ($T$) and polarization ($E$
and $B$) anisotropies, and in particular the identification of possible
local non-Gaussianity or statistical anisotropy signatures, or foreground
emissions, is thereby accessible with the same high precision in both
position and direction. The proposed algorithm may also be directly
applied in the study of beam asymmetry effects on the CMB temperature
and polarization data \cite{CMBPOLchallinor,SASwandelt,CMBPOLNg}.
But the generic results exposed for the directional correlation on
the sphere may actually find applications in many other fields beyond
cosmology \cite{SASbulow1,SASbulow2}.

In § \ref{sec:Directional-correlation}, we define the directional
correlation on the sphere as the scalar product of a signal and a
directional filter at all possible positions of the filter on the
sphere, and for all possible directions at each point. It is therefore
a function defined on the rotation group in three dimensions $SO(3)$.
By opposition, the standard correlation on the sphere considers a
fixed direction of the filter and is associated with a function on
the sphere $S^{2}$. We discuss the \emph{a priori} $\mathcal{O}(L^{5})$
asymptotic complexity for the computation of the directional correlation
through simple quadrature, where $2L$ stands for the square-root
of the number of sampling points on the sphere, also setting a band
limit $L$ for the signals and filters considered. The corresponding
naive computation of the directional correlation would take several
years on a single standard computer for fine samplings on the sphere
of order $L\simeq10^{3}$, and is thus inaccessible in terms of computation
times. We then recall that the spherical harmonics and the Wigner
$D$-functions form orthogonal bases in the spaces of square-integrable
functions on $S^{2}$ and $SO(3)$ respectively. We also introduce
a relation giving the directional correlation as the inverse Wigner
$D$-functions transform of the pointwise product between the signal
and filter scalar spherical harmonics coefficients. In § \ref{sec:Filter-steerability},
we recall the notion of steerable filters on the sphere, and give
the examples of the first and second Gaussian derivatives. We also
show that the directional correlation of arbitrary signals with steerable
filters is explicitly given in terms of the standard correlations
with their basis filters. We briefly introduce spin-weighted spherical
harmonics as orthonormal bases for the decomposition of spin functions
on the sphere and derive a natural relation giving the standard correlation
as a sum of inverse spin-weighted spherical harmonics transforms.
In § \ref{sec:Fast-algorithm}, we explicitly define our algorithm.
We first discuss the global algorithmic structure, as well as pixelization
choices. We then establish a recurrence relation which expresses spin-weighted
spherical harmonics as linear combinations of scalar spherical harmonics.
Consequently, the directional correlation of a scalar function with
a steerable filter may be uniquely expressed in terms of direct and
inverse scalar spherical harmonics transforms. The separation of variable
technique sets the asymptotic complexity of the scalar transform to
$\mathcal{O}(L^{3})$. The specific use of equi-angular pixelizations
allows to achieve the exactness of the algorithm, through a sampling
theorem introduced by Driscoll and Healy for their fast scalar spherical
harmonics transform. If the associated Legendre polynomials necessary
for the transform are pre-calculated, the asymptotic complexity of
the Driscoll and Healy transform on equi-angular grids drops to an
optimized $\mathcal{O}(L^{2}\log_{2}^{2}L)$. We discuss the numerical
implementations of the scalar spherical harmonics transforms on equi-angular
pixelizations (SpharmonicKit package) as well as on Hierarchical Equal-Area
iso-Latitude pixelizations (HEALPix package). A simple comparative
analysis of the corresponding memory requirements, computation times,
and numerical stability shows that both implementations of the directional
correlation algorithm are easily accessible on a standard computer
for resolutions well beyond $L\simeq10^{3}$. Finally, we briefly
recall an existing alternative $\mathcal{O}(L^{4})$ algorithm for
the directional correlation, which may also be optimized to an $\mathcal{O}(L^{3})$
asymptotic complexity through the use of steerable filters. In § \ref{sec:CMB-Scale-space-analysis},
we consider the algorithm in the perspective of the scale-space analysis
of the CMB, through the directional correlation of the temperature
$T$, and of the electric $E$ and magnetic $B$ polarization components.
We illustrate the performance of our algorithm through the explicit
computation of the wavelet coefficients of a simulated CMB temperature
map with the second Gaussian derivative wavelet. A comparison of the
results of the equi-angular and HEALPix implementations is proposed.
The good precision of the two implementations is confirmed with a
clear advantage though for the equi-angular approach which is theoretically
exact. The computation times of directional correlation of maps of
several megapixels are reduced from years to tens of seconds. We briefly
conclude in § \ref{sec:Conclusion}.

\section{Directional correlation}

\label{sec:Directional-correlation}

\subsection{Definition}

We define here explicitly the notions of directional and standard
correlations.

Let the function $F(\omega)$ and the filter $\Psi(\omega)$ be square-integrable
functions in $L^{2}(S^{2},d\Omega)$ on the unit sphere $S^{2}$.
The point $\omega$ is given in the spherical coordinates defined
in the right-handed Cartesian coordinate system $(o,o\hat{x},o\hat{y},o\hat{z})$
centered on the unit sphere as: $\omega=(\theta,\varphi)$. The angle
$\theta\in[0,\pi]$ is the polar angle, or co-latitude. The angle
$\varphi\in[0,2\pi[$ is the azimuthal angle, or longitude. The invariant
measure on the sphere reads $d\Omega=d\cos\theta d\varphi$. Recall
that an axisymmetric filter is by definition invariant under rotation
on itself. That is, when located at the North pole, an axisymmetric
filter is defined by a function $\Psi(\theta)$ independent of the
azimuthal angle $\varphi$. Any non-axisymmetric filter is said to
be directional, and is given as a general function $\Psi(\theta,\varphi)$
in $L^{2}(S^{2},d\Omega)$. The directional correlation is defined
as the scalar product between the function $F$ and an arbitrary filter
$\Psi$ rotated on itself in a direction $\chi\in[0,2\pi[$, and translated
at any point $\omega_{0}=(\theta_{0},\varphi_{0})$ on the sphere.
The Euler angles $(\varphi_{0},\theta_{0},\chi)$ associated with
a general rotation in three dimensions $\rho\in SO(3)$ represent
successive rotations by $\chi$ around $o\hat{z}$, $\theta_{0}$
around $o\hat{y}$, and $\varphi_{0}$ around $o\hat{z}$. These may
also be interpreted in the reverse order as successive rotations by
$\varphi_{0}$ around $o\hat{z}$, $\theta_{0}$ around $o\hat{y}'$,
and $\chi$ around $o\hat{z}''$, where the axes $o\hat{y}'\equiv o\hat{y}'(\varphi_{0})$
and $o\hat{z}''\equiv o\hat{z}''(\varphi_{0},\theta_{0})$ are respectively
obtained by the first and second rotations of the coordinate system
by $\varphi_{0}$ and $\theta_{0}$ \cite{CVbrink}. The translations
of the filter $\Psi$ at $\omega=(\theta_{0},\varphi_{0})$ are thus
associated with the two first Euler angles $\varphi_{0}$ around $o\hat{z}$,
$\theta_{0}$ around $o\hat{y}'$. The rotations of the filter on
itself in the plane tangent to the sphere at $\omega=(\theta_{0},\varphi_{0})$
are rotations around $o\hat{z}''$, associated with the third Euler
angle $\chi$. The affine transformations considered are therefore
implemented through the action on the function $\Psi$ of the operators
$R(\rho)$ associated with a rotation $\rho=(\varphi_{0},\theta_{0},\chi)$
in $SO(3)$. These operators read, in the first Euler angles interpretation,
as $R(\rho)=R^{\hat{z}}(\varphi_{0})R^{\hat{y}}(\theta_{0})R^{\hat{z}}(\chi)=R(\omega_{0})R^{\hat{z}}(\chi)$.
The rotation $R\Psi$ reads $[R(\rho)\Psi](\omega)=\Psi(R_{\rho}^{-1}\omega)$,
where $R_{\rho}=R_{\varphi_{0}}^{\hat{z}}R_{\theta_{0}}^{\hat{y}}R_{\chi}^{\hat{z}}=R_{\omega_{0}}R_{\chi}^{\hat{z}}$
stands for the three-dimensional rotation matrix associated with $\rho$,
acting on the coordinates $\omega=(\theta,\varphi)$. The result of
the directional correlation thus explicitly gives a square-integrable
function in $L^{2}(SO(3),d\rho)$ on the rotation group, for the invariant
measure $d\rho=d\cos\theta_{0}d\varphi_{0}d\chi$.

The directional correlation $\langle R\Psi\vert F\rangle$ of the
function $F$ by the filter $\Psi$ thus explicitly reads in $L^{2}(SO(3),d\rho)$
as:

\begin{equation}
\langle R\left(\rho\right)\Psi\vert F\rangle=\int_{S^{2}}d\Omega\,\Psi^{*}\left(R_{\rho}^{-1}\omega\right)F\left(\omega\right).\label{eq:dc-01}\end{equation}
The standard correlation $\langle R_{0}\Psi\vert F\rangle$ of $F$
by $\Psi$, is defined by the scalar product between the function
$F$ and the filter $\Psi$ translated at any point $\omega_{0}=(\theta_{0},\varphi_{0})$
on the sphere, but for a fixed direction, \emph{i.e.} a fixed value
$\chi=0$. The result of the standard correlation explicitly gives
a square-integrable function in $L^{2}(S^{2},d\Omega)$ on the sphere:\begin{equation}
\langle R\left(\omega_{0}\right)\Psi\vert F\rangle=\int_{S^{2}}d\Omega\,\Psi^{*}\left(R_{\omega_{0}}^{-1}\omega\right)F\left(\omega\right).\label{eq:dc-02}\end{equation}
 The notation $R_{0}$ simply identifies a three-dimensional rotation
with $\chi=0$. It distinguishes the standard correlation $\langle R_{0}\Psi\vert F\rangle$
from the directional correlation $\langle R\Psi\vert F\rangle$ when
the arguments are not specified. In the particular case of an axisymmetric
filter, there is no dependence of the correlation in the filter rotation
$\chi$. The directional correlation with an axisymmetric filter is
therefore strictly equivalent to the standard correlation.

Notice that if $\Psi$ is the specific dilation by $a\in\mathbb{R}_{+}^{*}$
of a wavelet on the sphere, $\Psi\rightarrow\Psi_{a}$, the directional
correlation identifies with the wavelet coefficient of the signal,
at the corresponding scale: $W_{\Psi}^{F}(\omega_{0},\chi,a)=\langle R(\rho)\Psi_{a}\vert F\rangle$,
for $\rho=(\varphi_{0},\theta_{0},\chi)$ \cite{SASwiaux1}. Let us
finally remark that the correlation is also generically called convolution
on the sphere \cite{SASwandelt}. We use a specific terminology to
avoid any confusion with other definitions \cite{SASdriscoll,SAShealy1}.

\subsection{A priori computation cost}

The numerical computation cost associated with a naive discretization,
or quadrature, of the relation (\ref{eq:dc-01}) for the directional
correlation of functions on the sphere may easily be estimated in
the following way.

First, let $N_{p}=(2L)^{2}$ represent the number of sampling points
$\omega$ in a given pixelization of the sphere. The quantity $2L$
represents the mean number of sampling points in the position variables
$\theta$ and $\varphi$. A simple extrapolation of the Nyquist-Shannon
theorem on the line intuitively associates $L$ with the band limit,
or maximum frequency, accessible in the {}``Fourier'' indices conjugate
to $\theta$ and $\varphi$ for the signals and filters considered.
To an equivalent sampling on $2L$ points in the direction $\chi$
also corresponds a band limit $L$ in the conjugate Fourier index.

Second, considering a simple discretization of the relation of directional
correlation, each integration on $\omega=(\theta,\varphi)$ on the
sphere, a scalar product, has an asymptotic complexity $\mathcal{O}(L^{2})$.
The directional correlation is associated with the calculation of
a scalar product for each discrete $\rho=(\varphi_{0},\theta_{0},\chi)$
on $SO(3)$. The corresponding naive asymptotic complexity is therefore
of order $\mathcal{O}(L^{5})$.

Third, we consider fine samplings of several megapixels on the sphere
associated with a band limit around $L\simeq10^{3}$. In particular,
present CMB experiments such as the WMAP mission provide all-sky maps
of around three megapixels. For such a fine sampling, the typical
computation times for $(2L)^{2}$ multiplications and $(2L)^{2}$
additions are of order of $0.03$ seconds on a standard $2.2$ GHz
Intel Pentium Xeon CPU, for double-precision numbers. We take this
value as a fair estimation of the computation time required for a
scalar product in (\ref{eq:dc-01}). Consequently, a unique $\mathcal{O}(L^{5})$
directional correlation would take several years at that band limit
$L\simeq10^{3}$ on a single standard computer. Moreover, depending
on the application, the directional correlation of multiple signals
might be required. Typically, thousands of simulated signals are to
be considered for a statistical (CMB) analysis.

In conclusion, the computation time for a simple $\mathcal{O}(L^{5})$
discretization of the relation for the directional correlation of
functions on the sphere is absolutely unaffordable for fine samplings
with a band limit around $L\simeq10^{3}$ in $\theta$, $\varphi$,
and $\chi$. This conclusion remains when the use of multiple computers
is envisaged. It is even strongly reinforced in the perspective of
a scale-space analysis on the sphere from finer pixelizations. In
particular, the future all-sky CMB maps of the Planck mission will
reach samplings of fifty megapixels\emph{, i.e.} $L\simeq4\times10^{3}$.

\subsection{Directional correlation as a Wigner transform}

We here discuss the decomposition of the directional correlation as
an inverse Wigner $D$-functions transform on $SO(3)$.

First, we recall the decomposition of functions on $S^{2}$ and $SO(3)$
in standard scalar spherical harmonics and Wigner $D$-functions respectively.
The standard scalar spherical harmonics $Y_{lm}(\omega)$, with $l\in\mathbb{N}$,
$m\in\mathbb{Z}$, and $|m|\leq l$, form an orthonormal basis for
the decomposition of functions in $L^{2}(S^{2},d\Omega)$ on the sphere,
with $\omega=(\theta,\varphi)$ \cite{SASvarshalovich}. They are
explicitly given in a factorized form in terms of the associated Legendre
polynomials $P_{l}^{m}(\cos\theta)$ and the complex exponentials
$e^{im\varphi}$ as \begin{equation}
Y_{lm}(\theta,\varphi)=\left[\frac{2l+1}{4\pi}\frac{\left(l-m\right)!}{\left(l+m\right)!}\right]^{1/2}P_{l}^{m}\left(\cos\theta\right)e^{im\varphi}.\label{eq:sv-01}\end{equation}
This corresponds to the choice of Condon-Shortley phase $(-1)^{m}$
for the spherical harmonics, ensuring the relation $(-1)^{m}Y_{lm}^{*}(\omega)=Y_{l(-m)}(\omega)$.
This phase is here included in the definition of the associated Legendre
polynomials \cite{SASabramowitz,SASvarshalovich}. Another convention
\cite{CVbrink} explicitly transfers it to the spherical harmonics.
The orthonormality and completeness relations respectively read:\begin{equation}
\int_{S^{2}}d\Omega\, Y_{lm}^{*}\left(\omega\right)Y_{l'm'}\left(\omega\right)=\delta_{ll'}\delta_{mm'},\label{eq:sv-02a}\end{equation}
and \begin{equation}
\sum_{l\in\mathbb{N}}\sum_{|m|\leq l}Y_{lm}^{*}\left(\omega'\right)Y_{lm}\left(\omega\right)=\delta\left(\omega'-\omega\right),\label{eq:sv-02b}\end{equation}
with $\delta(\omega'-\omega)=\delta(\cos\theta'-\cos\theta)\delta(\varphi'-\varphi)$.
Any function $G(\omega)$ on the sphere is thus uniquely given as
a linear combination of scalar spherical harmonics (inverse transform):
\begin{equation}
G\left(\omega\right)=\sum_{l\in\mathbb{N}}\sum_{|m|\leq l}\widehat{G}_{lm}Y_{lm}\left(\omega\right),\label{eq:sv-03}\end{equation}
for the scalar spherical harmonics coefficients (direct transform)
\begin{equation}
\widehat{G}_{lm}=\int_{S^{2}}d\Omega\, Y_{lm}^{*}\left(\omega\right)G\left(\omega\right),\label{eq:sv-04}\end{equation}
with $\vert m\vert\leq l$.

The Wigner $D$-functions $D_{mn}^{l}(\rho)$, for $\rho=(\varphi,\theta,\chi)\in SO(3)$,
and with $l\in\mathbb{N}$, $m,n\in\mathbb{Z}$, and $|m|,|n|\leq l$,
are the matrix elements of the irreducible unitary representations
of weight $l$ of the rotation group $SO(3)$, in $L^{2}(SO(3),d\rho)$.
By the Peter-Weyl theorem on compact groups, the matrix elements $D_{mn}^{l*}$
also form an orthogonal basis in $L^{2}(SO(3),d\rho)$ \cite{SASvarshalovich}.
They are explicitly given in a factorized form in terms of the real
Wigner $d$-functions $d_{mn}^{l}(\theta)$ and the complex exponentials,
$e^{-im\varphi}$ and $e^{-in\chi}$ , as\begin{equation}
D_{mn}^{l}\left(\varphi,\theta,\chi\right)=e^{-im\varphi}d_{mn}^{l}\left(\theta\right)e^{-in\chi}.\label{eq:sv-05}\end{equation}
The Wigner $d$-functions read\begin{eqnarray}
d_{mn}^{l}\left(\theta\right) & = & \sum_{t=C_{1}}^{C_{2}}\frac{\left(-1\right)^{t}\left[\left(l+m\right)!\left(l-m\right)!\left(l+n\right)!\left(l-n\right)!\right]^{1/2}}{\left(l+m-t\right)!\left(l-n-t\right)!t!\left(t+n-m\right)!}\nonumber \\
 &  & \left(\cos\theta/2\right)^{2l+m-n-2t}\left(\sin\theta/2\right)^{2t+n-m},\label{eq:sv-06}\end{eqnarray}
 with the summation bounds $C_{1}=\max(0,m-n)$ and $C_{2}=\min(l+m,l-n)$
defined to consider only factorials of positive integers. They satisfy
various symmetry properties on their indices \cite{SASvarshalovich}.
The orthogonality and completeness relations of the Wigner $D$-functions
respectively read: \begin{equation}
\int_{SO(3)}d\rho\, D_{mn}^{l}\left(\rho\right)D_{m'n'}^{l'*}\left(\rho\right)=\frac{8\pi^{2}}{2l+1}\delta_{ll'}\delta_{mm'}\delta_{nn'},\label{eq:sv-07a}\end{equation}
and \begin{equation}
\sum_{l\in\mathbb{N}}\frac{2l+1}{8\pi^{2}}\sum_{|m|,|n|\leq l}D_{mn}^{l}\left(\rho'\right)D_{mn}^{l*}\left(\rho\right)=\delta\left(\rho'-\rho\right),\label{eq:sv-07b}\end{equation}
with $\delta(\rho'-\rho)=\delta(\varphi'-\varphi)\delta(\cos\theta'-\cos\theta)\delta(\chi'-\chi)$.
Any function $G(\rho)$ in $L^{2}(SO(3),d\rho)$ is thus uniquely
given as a linear combination of Wigner $D$-functions (inverse transform):\begin{equation}
G\left(\rho\right)=\sum_{l\in\mathbb{N}}\frac{2l+1}{8\pi^{2}}\sum_{|m|,|n|\leq l}\widehat{G}_{mn}^{l}D_{mn}^{l*}\left(\rho\right),\label{eq:sv-08}\end{equation}
for the Wigner $D$-functions coefficients (direct transform) \begin{equation}
\widehat{G}_{mn}^{l}=\int_{SO(3)}d\rho\, D_{mn}^{l}\left(\rho\right)G\left(\rho\right),\label{eq:sv-09}\end{equation}
with $\vert m\vert,\vert n\vert\leq l$.

Second, we expose the decomposition of the directional correlation
in Wigner $D$-functions. The Wigner $D$-functions coefficients $\widehat{\langle R\Psi\vert F\rangle}_{mn}^{l}$
of the directional correlation $\langle R(\rho)\Psi\vert F\rangle$
are given as the pointwise product of the scalar spherical harmonics
coefficients $\widehat{F}_{lm}$ and $\widehat{\Psi}_{ln}^{*}$. The
following directional correlation relation indeed holds:\begin{equation}
\langle R\left(\rho\right)\Psi\vert F\rangle=\sum_{l\in\mathbb{N}}\frac{2l+1}{8\pi^{2}}\sum_{|m|,|n|\leq l}\widehat{\langle R\Psi\vert F\rangle}_{mn}^{l}D_{mn}^{l*}\left(\rho\right),\label{eq:sv-10}\end{equation}
with the Wigner $D$-functions coefficients on $SO(3)$ $\widehat{\langle R\Psi\vert F\rangle}_{mn}^{l}$
given as\begin{equation}
\widehat{\langle R\Psi\vert F\rangle}_{mn}^{l}=\frac{8\pi^{2}}{2l+1}\widehat{\Psi}_{ln}^{*}\widehat{F}_{lm}.\label{eq:sv-11}\end{equation}
The derivation of this result goes as follows. The orthonormality
of scalar spherical harmonics implies the following Plancherel relation
$\langle R\Psi\vert F\rangle=\sum_{l\in\mathbb{N}}\sum_{|m|\leq l}\widehat{R\Psi}_{lm}^{*}\widehat{F}_{lm}$.
The action of the operator $R(\rho)$ on a function $G(\omega)$ in
$L^{2}(S^{2},d\Omega)$ on the sphere reads in terms of its scalar
spherical harmonics coefficients and the Wigner $D$-functions as:
$\widehat{[R(\rho)G]}_{lm}=\sum_{|n|\leq l}D_{mn}^{l}(\rho)\widehat{G}_{ln}.$
Inserting this last relation for $\Psi$ in the former Plancherel
relation finally gives the result. In the particular case of an axisymmetric
filter $\Psi(\theta)$, the above relation reduces to the following
standard correlation relation: $\widehat{\langle R_{0}\Psi\vert F\rangle}_{lm}=2\pi\widehat{\Psi}_{l}^{*}\widehat{F}_{lm}$,
where $\widehat{\Psi}_{l}$ stands for the Legendre coefficient of
the filter.

\section{Filter steerability}

\label{sec:Filter-steerability}

\subsection{Definition}

We here recall the notion of filter steerability on the sphere \cite{SASwiaux1,STfreeman,STsimoncelli}
and give explicit examples. A directional filter $\Psi$ in $L^{2}(S^{2},d\Omega)$
on the sphere is steerable if any rotation by $\chi\in[0,2\pi[$ of
the filter around itself $R^{\hat{z}}(\chi)\Psi$ may be expressed
as a linear combination of a finite number of basis filters $\Psi_{m}$:\begin{equation}
\left[R^{\hat{z}}\left(\chi\right)\Psi\right]\left(\omega\right)=\sum_{m=1}^{M}k_{m}\left(\chi\right)\Psi_{m}\left(\omega\right).\label{eq:fs-01}\end{equation}
 The weights $k_{m}(\chi)$, with $1\leq m\leq M$, and $M\in\mathbb{N}$,
are called interpolation functions. In particular cases, the basis
filters may be specific rotations by angles $\chi_{m}$ of the original
filter: $\Psi_{m}=R^{\hat{z}}(\chi_{m})\Psi$. Steerable filters have
a non-zero angular width in the azimuthal angle $\varphi$ which makes
them sensitive to a whole range of directions and enables them to
satisfy the relation (\ref{eq:fs-01}). In the spherical harmonics
space, this non-zero angular width corresponds to an azimuthal angular
band limit $N$ in the index $n$ associated with the azimuthal variable
$\varphi$: $\widehat{\Psi}_{ln}=0$ for $\vert n\vert\geq N$.

The inverse stereographic projection on the sphere of the $N_{d}^{th}$
derivative in direction $\hat{x}$ of radial functions on the plane
(tangent at the North pole) are steerable filters. They may be rotated
in terms of $M=N_{d}+1$ basis filters, and are band-limited in $\varphi$
at $N=N_{d}+1$. We give here the explicit examples of the normalized
first and second Gaussian derivatives. A first derivative has a band
limit $N=2$, and only contains the indices $n=\{\pm1\}$. It may
be rotated in terms of two specific rotations at $\chi=0$ and $\chi=\pi/2$,
corresponding to the inverse projection of the first derivatives in
directions $\hat{x}$ and $\hat{y}$, $\Psi^{\partial_{\hat{x}}}$
and $\Psi^{\partial_{\hat{y}}}$ respectively:\begin{equation}
\left[R^{\hat{z}}\left(\chi\right)\Psi^{\partial_{\hat{x}}}\right]\left(\omega\right)=\Psi^{\partial_{\hat{x}}}\left(\omega\right)\cos\chi+\Psi^{\partial_{\hat{y}}}\left(\omega\right)\sin\chi.\label{eq:fs-02}\end{equation}
The normalized first derivatives of a Gaussian in directions $\hat{x}$
and $\hat{y}$ read:\begin{eqnarray}
\Psi^{\partial_{\hat{x}}(gauss)}\left(\theta,\varphi\right) & = & \sqrt{\frac{8}{\pi}}\left(1+\tan^{2}\frac{\theta}{2}\right)e^{-2\tan^{2}(\theta/2)}\tan\frac{\theta}{2}\cos\varphi\nonumber \\
\Psi^{\partial_{\hat{y}}(gauss)}\left(\theta,\varphi\right) & = & \sqrt{\frac{8}{\pi}}\left(1+\tan^{2}\frac{\theta}{2}\right)e^{-2\tan^{2}(\theta/2)}\tan\frac{\theta}{2}\sin\varphi.\nonumber \\
\label{eq:fs-03}\end{eqnarray}
A second derivative has a band limit $N=3$, and contains the frequencies
$n=\{0,\pm2\}$. It may be rotated in terms of three basis filters.
It reads indeed in terms of the inverse projection of the second derivatives
in directions $\hat{x}$ and $\hat{y}$, $\Psi^{\partial_{\hat{x}}^{2}}$
and $\Psi^{\partial_{\hat{y}}^{2}}$ respectively, and the cross derivative
$\Psi^{\partial_{\hat{x}}\partial_{\hat{y}}}$ as:\begin{eqnarray}
\left[R^{\hat{z}}\left(\chi\right)\Psi^{\partial_{\hat{x}}^{2}}\right]\left(\omega\right) & = & \Psi^{\partial_{\hat{x}}^{2}}\left(\omega\right)\cos^{2}\chi+\Psi^{\partial_{\hat{y}}^{2}}\left(\omega\right)\sin^{2}\chi\nonumber \\
 &  & +\Psi^{\partial_{\hat{x}}\partial_{\hat{y}}}\left(\omega\right)\sin2\chi.\label{eq:fs-04}\end{eqnarray}
The correctly normalized second derivatives of a Gaussian in directions
$\hat{x}$ and $\hat{y}$ read:\begin{eqnarray}
\Psi^{\partial_{\hat{x}}^{2}(gauss)}\left(\theta,\varphi\right) & = & \sqrt{\frac{4}{3\pi}}\left(1+\tan^{2}\frac{\theta}{2}\right)e^{-2\tan^{2}(\theta/2)}\nonumber \\
 &  & \left(1-4\tan^{2}\frac{\theta}{2}\cos^{2}\varphi\right)\nonumber \\
\Psi^{\partial_{\hat{y}}^{2}(gauss)}\left(\theta,\varphi\right) & = & \sqrt{\frac{4}{3\pi}}\left(1+\tan^{2}\frac{\theta}{2}\right)e^{-2\tan^{2}(\theta/2)}\nonumber \\
 &  & \left(1-4\tan^{2}\frac{\theta}{2}\sin^{2}\varphi\right)\nonumber \\
\Psi^{\partial_{\hat{x}}\partial_{\hat{y}}(gauss)}\left(\theta,\varphi\right) & = & -\frac{4}{\sqrt{3\pi}}\left(1+\tan^{2}\frac{\theta}{2}\right)e^{-2\tan^{2}(\theta/2)}\nonumber \\
 &  & \left(\tan^{2}\frac{\theta}{2}\sin2\varphi\right).\label{eq:fs-05}\end{eqnarray}

\subsection{From directional to standard correlation}

We here express the directional correlation of arbitrary signals with
steerable filters in terms of standard correlations.

The relation of steerability (\ref{eq:fs-01}) is obviously preserved
by linear filtering. The directional correlation of a signal $F$
by a steerable filter $\Psi$ therefore satisfies the same steerability
relation as the filter itself:\begin{equation}
\langle R\left(\rho\right)\Psi\vert F\rangle=\sum_{m=1}^{M}k_{m}\left(\chi\right)\langle R\left(\omega_{0}\right)\Psi_{m}\vert F\rangle,\label{eq:fs-06}\end{equation}
for $\rho=(\varphi_{0},\theta_{0},\chi)$, and $\omega_{0}=(\varphi_{0},\theta_{0})$.
The steerability therefore enables the computation of the directional
correlation of a signal $F$ with a steerable filter $\Psi$ as a
linear combination with known weights $k_{m}(\chi)$, of $M$ standard
correlations with the basis filters $\Psi_{m}$. For steerable filters
with $M<<L$, this quantity $M$ disappears from asymptotic complexities
calculations. In conclusion, the asymptotic complexity of the directional
correlation with a steerable filter is reduced to the asymptotic complexity
of a standard correlation, to which must be added the $\mathcal{O}(L^{3})$
operations required for the explicit computation of the directional
correlation through the relation (\ref{eq:fs-06}). Already notice
that the \emph{a priori} computation cost associated with the discretization
of the directional correlation integral (\ref{eq:dc-01}) with a steerable
filter is lowered from $\mathcal{O}(L^{5})$ to $\mathcal{O}(L^{4})$
through the calculation of standard correlation integrals (\ref{eq:dc-02})
with the corresponding basis filters.

\subsection{Standard correlation as a spin-weighted spherical harmonics transform}

We here study the decomposition of the standard correlation as a sum
of inverse spin-weighted spherical harmonics transforms on the sphere.

Let us first recall the notion of spin $n$ function in $L^{2}(S^{2},d\Omega)$
on the sphere and the related decomposition in spin-weighted spherical
harmonics of spin $n$. As defined in § \ref{sec:Directional-correlation},
in the coordinate frame $(o,o\hat{x},o\hat{y},o\hat{z})$, the Euler
angles $(\varphi,\theta,\chi)$ associated with a general rotation
in three dimensions may be interpreted as successive rotations by
$\varphi\in[0,2\pi[$ around $o\hat{z}$, $\theta\in[0,\pi]$ around
$o\hat{y}'(\varphi)$, and $\chi\in[0,2\pi[$ around $o\hat{z}''(\varphi,\theta)$.
The local rotations of the basis vectors in the plane tangent to the
sphere at $\omega=(\theta,\varphi)$ are rotations around $o\hat{z}''$,
associated with the third Euler angle $\chi$. Spin $n$ functions
on the sphere ${}_{n}G(\omega)$, with $n\in\mathbb{Z}$, are defined
relatively to their behaviour under the corresponding right-handed
rotations by $\chi_{0}$ as \cite{SASnewman,SASgoldberg,SAScarmeli}:\begin{equation}
{}_{n}G'\left(\omega\right)=e^{-in\chi_{0}}\,\,{}_{n}G\left(\omega\right).\label{eq:fs-07}\end{equation}
 We emphasize that the rotations considered are local transformations
on the sphere around the axis $o\hat{z}''\equiv o\hat{z}''(\varphi,\theta)$,
affecting the coordinate $\chi$ in the tangent plane independently
at each point $\omega=(\theta,\varphi)$, according to $\chi'=\chi-\chi_{0}$.
They are to be clearly distinguished from the global rotations $R_{\chi}^{\hat{z}}$
associated with the alternative Euler angles interpretation, which
affect the coordinates of the points $\omega=(\theta,\varphi)$ on
the sphere. Our sign convention in the exponential is coherent with
the definition (\ref{eq:fs-08}) here below for the spin-weighted
spherical harmonics of spin $n$. It is opposite to the original definition
\cite{SASnewman}, while equivalent to recent notations used in the
context of the CMB analysis \cite{CMBzaldarriaga,CMBPOLchallinor}.
Recalling the factorized form (\ref{eq:sv-05}), spin functions are
equivalently defined as the evaluation at $\chi=0$ of any function
in $L^{2}(SO(3),d\rho)$ resulting from an expansion for fixed index
$n$ in the Wigner $D$-functions $D_{mn}^{l}(\varphi,\theta,\chi)$.
The functions $D_{mn}^{l}(\varphi,\theta,0)$ or $D_{m(-n)}^{l*}(\varphi,\theta,0)$
thus naturally define for each $n$ an orthogonal basis for the expansion
of spin $n$ functions in $L^{2}(S^{2},d\Omega)$ on the sphere. Their
normalization in $L^{2}(S^{2},d\Omega)$ defines the spin-weighted
spherical harmonics of spin $n$:\begin{equation}
{}_{n}Y_{lm}\left(\theta,\varphi\right)=\left(-1\right)^{n}\sqrt{\frac{2l+1}{4\pi}}D_{m(-n)}^{l*}\left(\varphi,\theta,0\right),\label{eq:fs-08}\end{equation}
with $l\in\mathbb{N}$, $l\geq\vert n\vert$, and $m\in\mathbb{Z}$,
$|m|\leq l$. They are thus explicitly given in a factorized form
in terms of the real Wigner $d$-functions $d_{mn}^{l}(\theta)$ and
the complex exponentials $e^{im\varphi}$ as \begin{equation}
{}_{n}Y_{lm}\left(\theta,\varphi\right)=\left(-1\right)^{n}\sqrt{\frac{2l+1}{4\pi}}d_{m(-n)}^{l}\left(\theta\right)e^{im\varphi}.\label{eq:fs-09}\end{equation}
In particular, the symmetry properties of the Wigner $d$-functions
imply the generalized relation $(-1)^{n+m}\,\,{}_{n}Y_{lm}^{*}(\omega)=\,\,{}_{-n}Y_{l(-m)}(\omega)$.
The spin $0$ spherical harmonics explicitly identify with the standard
scalar spherical harmonics: ${}_{0}Y_{lm}(\omega)=Y_{lm}(\omega)$,
through the relation $d_{m0}^{l}(\theta)=[\frac{(l-m)!}{(l+m)!}]^{1/2}P_{l}^{m}(\cos\theta)$.
The orthonormality and completeness relations respectively read from
relations (\ref{eq:sv-07a}) and (\ref{eq:sv-07b}), as\begin{equation}
\int_{S^{2}}d\Omega\,\,{}_{n}Y_{lm}^{*}\left(\omega\right)\,\,{}_{n}Y_{l'm'}\left(\omega\right)=\delta_{ll'}\delta_{mm'},\label{eq:fs-10a}\end{equation}
and \begin{equation}
\sum_{l\in\mathbb{N}}\sum_{|m|\leq l}\,\,{}_{n}Y_{lm}^{*}\left(\omega'\right)\,\,{}_{n}Y_{lm}\left(\omega\right)=\delta\left(\omega'-\omega\right),\label{eq:fs-10b}\end{equation}
with $\delta(\omega'-\omega)=\delta(\cos\theta'-\cos\theta)\delta(\varphi'-\varphi)$.
Any spin $n$ function ${}_{n}G(\omega)$ on the sphere is thus uniquely
given as a linear combination of spin $n$ spherical harmonics (inverse
transform): \begin{equation}
{}_{n}G\left(\omega\right)=\sum_{l\in\mathbb{N}}\sum_{|m|\leq l}\,\,{}_{n}\widehat{G}_{lm}\,\,{}_{n}Y_{lm}\left(\omega\right),\label{eq:fs-11}\end{equation}
for the spin-weighted spherical harmonics coefficients (direct transform)
\begin{equation}
{}_{n}\widehat{G}_{lm}=\int_{S^{2}}d\Omega\,\,{}_{n}Y_{lm}^{*}\left(\omega\right)G\left(\omega\right),\label{eq:fs-12}\end{equation}
with $l\geq\vert n\vert$, and $\vert m\vert\leq l$.

The directional correlation at $\chi=0$ identifies with the standard
correlation. From the above definitions, the relation (\ref{eq:sv-10})
therefore leads to the following standard correlation relation on
the sphere, expressed as a sum of inverse spin-weighted spherical
harmonics transforms:\begin{equation}
\langle R\left(\omega_{0}\right)\Psi\vert F\rangle=\sum_{n\in\mathbb{Z}}\left[\sum_{l\geq\vert n\vert}\sum_{|m|\leq l}\widehat{\langle R_{0}\Psi\vert F\rangle}_{lmn}\,\,{}_{n}Y_{lm}\left(\omega_{0}\right)\right],\label{eq:fs-13}\end{equation}
with coefficients $\widehat{\langle R_{0}\Psi\vert F\rangle}_{lmn}$
explicitly defined as the pointwise products\begin{equation}
\widehat{\langle R_{0}\Psi\vert F\rangle}_{lmn}=\left(-1\right)^{n}\sqrt{\frac{4\pi}{2l+1}}\widehat{\Psi}_{l(-n)}^{*}\widehat{F}_{lm}.\label{eq:fs-14}\end{equation}
For each $n$, these coefficients may be understood as spin-weighted
spherical harmonics coefficients, but to be clearly distinguished
from the spin-weighted spherical harmonics coefficients ${}_{n}\widehat{\langle R_{0}\Psi\vert F\rangle}_{lm}$
of the standard correlation, which do not find a pointwise product
expression. Considering a real filter also implies the further simplification:
$(-1)^{n}\widehat{\Psi}_{l(-n)}^{*}=\widehat{\Psi}_{ln}$.

\section{Fast algorithm}

\label{sec:Fast-algorithm}

\subsection{Algorithm and pixelization}

In the following, we define the global structure of our algorithm
and discuss pixelization choices.

In terms of the relations (\ref{eq:fs-13}) and (\ref{eq:fs-14}),
the computation of the standard correlation of a signal $F$ by a
steerable filter $\Psi$ may be performed with the following global
structure: a direct scalar spherical harmonics transform of the signal
and the filter, $\widehat{\Psi}_{ln}$ and $\widehat{F}_{lm}$, a
correlation in spherical harmonics space through a pointwise product,
and a sum of inverse spin-weighted spherical harmonics transforms
to obtain $\langle R_{0}\Psi\vert F\rangle$.

We consider band-limited signals $F$ at some band limit $L$ on the
sphere $S^{2}$, that is by definition $\widehat{F}_{lm}=0$ for $l\geq L$.
From (\ref{eq:fs-14}), the standard correlation of a band-limited
signal $F$ with a band limit $L$ by an band-limited filter $\Psi$
on the sphere is thus also band-limited, at the same band limit: $\widehat{\langle R_{0}\Psi\vert F\rangle}_{lmn}=0$
for $l\geq L$. In order to achieve the limit frequency $L$, an extrapolation
of the Nyquist-Shannon theorem on the line typically requires pixelizations
with at least $2L$ sampling points in the polar angle $\theta\in[0,\pi]$
and the azimuthal angle $\varphi\in[0,2\pi[$ on the sphere. In the
context of the signal processing of the CMB maps, many pixelization
schemes have been considered. We only quote here the HEALPix scheme
(Hierarchical Equal Area iso-Latitude Pixelization) \cite{SASgorski}
notably used for the WMAP and Planck experiments, and the GLESP pixelization
(Gauss-Legendre Sky Pixelization) \cite{SASdoroshkevich1,SASdoroshkevich2}.
In particular, on a $2L\times2L$ equi-angular grid, a sampling result
on the sphere states that the direct scalar and spin-weighted spherical
harmonics coefficients of a band-limited function on the sphere may
be computed exactly up to a band limit $L$ as a finite weighted sum,
\emph{i.e.} a quadrature, of the sampled values of that function \cite{SASdriscoll,SASkostelec1}.
The weights are defined from the structure of the Legendre polynomials
$P_{l}(\cos\theta)$ on $[0,\pi]$. They are functions of $\theta$,
but are independent of the azimuthal angle $\varphi$ and of the spin
$n$ considered. The theoretical exactness of computation represents
an advantage of equi-angular grids relative to other samplings. Inverse
scalar and spin-weighted spherical harmonics transforms of band-limited
functions can obviously be evaluated exactly on any grid as they are
explicitly defined as finite sums from relations (\ref{eq:sv-03})
and (\ref{eq:fs-11}) with $l<L$.

\subsection{From spin-weighted to scalar spherical harmonics}

We propose here an original expression of spin-weighted spherical
harmonics as simple linear combinations of scalar spherical harmonics.

The spin $n$ spherical harmonics may be related to spin $n\pm1$
spherical harmonics through the action of spin raising and lowering
operators \cite{SASnewman,SASgoldberg}. The action of the spin raising
$\eth$ and lowering $\bar{\eth}$ operators on a spin $n$ function
${}_{n}G$, giving a spin $n+1$ and $n-1$ function respectively,
is defined as \begin{equation}
\left[\eth\,\,{}_{n}G\right]\left(\theta,\varphi\right)=\left[-\sin^{n}\theta\left(\frac{\partial}{\partial\theta}+\frac{i}{\sin\theta}\frac{\partial}{\partial\varphi}\right)\sin^{-n}\theta\,\,{}_{n}G\right]\left(\theta,\varphi\right),\label{eq:fs-15}\end{equation}
 and \begin{equation}
\left[\bar{\eth}\,\,{}_{n}G\right]\left(\theta,\varphi\right)=\left[-\sin^{-n}\theta\left(\frac{\partial}{\partial\theta}-\frac{i}{\sin\theta}\frac{\partial}{\partial\varphi}\right)\sin^{n}\theta\,\,{}_{n}G\right]\left(\theta,\varphi\right),\label{eq:fs-16}\end{equation}
with, under rotation by $\chi_{0}$ in the tangent plane at $\omega=(\theta,\varphi)$:
$[\eth\,\,{}_{n}G]'(\omega)=e^{-i(n+1)\chi_{0}}[\eth\,\,{}_{n}G](\omega)$
and $[\bar{\eth}\,\,{}_{n}G]'(\omega)=e^{-i(n-1)\chi_{0}}[\bar{\eth}\,\,{}_{n}G](\omega)$.
In these terms, the spin-weighted spherical harmonics of spin $n$
are related to spin-weighted spherical harmonics of spin $n+1$ and
$n-1$ through the following relations:\begin{equation}
\left[\eth\,\,{}_{n}Y_{lm}\right]\left(\omega\right)=\left[\left(l-n\right)\left(l+n+1\right)\right]^{1/2}\,\,{}_{n+1}Y_{lm}\left(\omega\right),\label{eq:fs-17}\end{equation}
and\begin{equation}
\left[\bar{\eth}\,\,{}_{n}Y_{lm}\right]\left(\omega\right)=-\left[\left(l+n\right)\left(l-n+1\right)\right]^{1/2}\,\,{}_{n-1}Y_{lm}\left(\omega\right),\label{eq:fs-18}\end{equation}
also implying\begin{equation}
\left[\bar{\eth}\eth\,\,{}_{n}Y_{lm}\right]\left(\omega\right)=-\left(l-n\right)\left(l+n+1\right)\,\,{}_{n}Y_{lm}\left(\omega\right).\label{eq:fs-19}\end{equation}
The corresponding direct relation between the spin-weighted spherical
harmonics of spin $n$ and scalar spherical harmonics reads:\begin{equation}
{}_{n}Y_{lm}\left(\omega\right)=\left[\frac{\left(l-n\right)!}{\left(l+n\right)!}\right]^{1/2}\left[\eth^{n}\, Y_{lm}\right]\left(\omega\right),\label{eq:fs-20}\end{equation}
for $0\leq n\leq l$, and \begin{equation}
{}_{n}Y_{lm}\left(\omega\right)=\left[\frac{\left(l+n\right)!}{\left(l-n\right)!}\right]^{1/2}\left(-1\right)^{n}\left[\bar{\eth}^{-n}\, Y_{lm}\right]\left(\omega\right),\label{eq:fs-21}\end{equation}
for $-l\leq n\leq0$.

Beyond these standard relations, we establish a recurrence relation
free of derivatives for spin-weighted spherical harmonics as follows.
Derivative relations satisfied by the Wigner $D$-functions \cite{SASvarshalovich}
allow to exchange the derivative in $\theta$ in (\ref{eq:fs-17})
and (\ref{eq:fs-18}) for simple linear combinations. This gives the
spin $n$ spherical harmonics ${}_{n}Y_{lm}$ as a linear combination
of the spin $n\mp1$ spherical harmonics ${}_{n\mp1}Y_{lm}$ and ${}_{n\mp1}Y_{(l-1)m}$:\begin{eqnarray}
{}_{n}Y_{lm}\left(\theta,\varphi\right) & = & \left(\frac{l\mp n+1}{l\pm n}\right)^{1/2}\frac{m\mp l\cos\theta}{l\sin\theta}\,\,{}_{n\mp1}Y_{lm}\left(\theta,\varphi\right)\pm\left[\frac{2l+1}{2l-1}\right.\nonumber \\
 &  & \left.\frac{\left(l\pm n-1\right)\left(l^{2}-m^{2}\right)}{l\pm n}\right]^{1/2}\frac{1}{l\sin\theta}\,\,{}_{n\mp1}Y_{(l-1)m}\left(\theta,\varphi\right),\nonumber \\
\label{eq:fs-22}\end{eqnarray}
 under the convention that ${}_{n}Y_{lm}$ is defined to be zero for
$l<\max(\vert m\vert,\vert n\vert)$. Grouping terms through their
$1/\sin\theta$ or $\cot\theta$ pre-factors on the sphere, one gets
in compact form the following $2$-terms relation:\begin{equation}
{}_{n}Y_{lm}\left(\theta,\varphi\right)=\left(\frac{\frac{m}{l}\alpha_{(ln)}^{\pm}\pm\beta_{(lmn)}^{\pm}S^{-1}}{\sin\theta}\mp\alpha_{(ln)}^{\pm}\cot\theta\right)\,\,{}_{n\mp1}Y_{lm}\left(\theta,\varphi\right),\label{eq:fs-23}\end{equation}
 with $\alpha_{(ln)}^{\pm}=(\frac{l\mp n+1}{l\pm n})^{1/2}$, and
$\beta_{(lmn)}^{\pm}=\frac{1}{l}[\frac{2l+1}{2l-1}\frac{(l\pm n-1)(l^{2}-m^{2})}{l\pm n}]^{1/2}$.
The operators $S^{\pm1}:l\rightarrow l\pm1$ define a one-unit shift
in the index $l$: $[S^{-1}\,\,{}_{n\mp1}Y_{lm}](\omega)=\,\,{}_{n\mp1}Y_{(l-1)m}(\omega)$.
By a simple $\vert n\vert$-steps recurrence, spin-weighted spherical
harmonics may therefore by expressed as a $2^{\vert n\vert}$-terms
linear combination of scalar spherical harmonics. The same relation
between spin-weighted and scalar spherical harmonics may also be obtained
from relations (\ref{eq:fs-20}) and (\ref{eq:fs-21}) and direct
derivative relations for the associated Legendre polynomials \cite{SASabramowitz}.

\subsection{Detailed algorithmic structure from scalar spherical harmonics transforms}

\label{sub:Detailed-algorithmic-structure}

In the following, we discuss the detailed structure of the algorithm,
which may essentially be decomposed in a combination of scalar spherical
harmonics transforms, and study the corresponding asymptotic complexity.

From the recurrence (\ref{eq:fs-23}), the inverse spin-weighted transforms
of $\widehat{\langle R_{0}\Psi\vert F\rangle}_{lmn}$ at each $n$
required in the relation (\ref{eq:fs-13}) for the standard correlation
may be decomposed as a linear combination of inverse scalar spherical
harmonics transforms. Hence, the complete standard correlation of
a band-limited signal with a band-limited steerable filter essentially
relies on direct and inverse scalar spherical harmonics transforms.

The \emph{a priori} complexity associated with the naive computation
of the direct scalar transform integral (\ref{eq:sv-04}) in $(\theta,\varphi)$
on the sphere through simple quadrature, for all $(l,m)$ with $\vert m\vert\leq l<L$,
or with the sum on $(l,m)$ for all discrete points $(\theta,\varphi)$
in the inverse scalar transform (\ref{eq:sv-03}) is naturally of
order $\mathcal{O}(L^{4})$. The separation of variables (\ref{eq:sv-01})
in the scalar spherical harmonics into the associated Legendre polynomials
$P_{l}^{m}(\cos\theta)$ and the complex exponentials $e^{im\varphi}$
allows to compute direct and inverse scalar spherical harmonics transforms
as successive transforms in each of the variables $\theta$ and $\varphi$
in $\mathcal{O}(L^{3})$ operations. The application of the concept
of separation of variables on the sphere only requires that the sampling
in the polar angle $\theta$ be independent of the azimuthal angle
$\varphi$. This criterion is met for a large variety of pixelizations.
We will consider the corresponding stable numerical implementations
on HEALPix pixelizations \cite{SASgorski}%
\footnote{See healpix.jpl.nasa.gov for documentation and algorithm implementation
(HEALPix2.1 software).%
}. For a $2L\times2L$ equi-angular grid in $(\theta,\varphi)$ on
the sphere, a further optimized algorithm exists which is due to Driscoll
and Healy \cite{SASdriscoll}. It explicitly takes advantage of a
recurrence relation in $l$ on the associated Legendre polynomials
$P_{l}^{m}(\cos\theta)$ to compute the associated Legendre transforms
in $\mathcal{O}(L\log_{2}^{2}L)$ operations for each $m$. The Fourier
transforms in $e^{im\varphi}$ are computed in $\mathcal{O}(L\log_{2}L)$
operations for each $\theta$ through standard Cooley-Tukey fast Fourier
transforms. In these terms, the direct and inverse scalar spherical
harmonics transforms of a band-limited function of band limit $L$
on the sphere are computed in $\mathcal{O}(L^{2}\log_{2}^{2}L)$ operations.
Moreover, contrarily to the HEALPix implementation, the algorithm
is theoretically exact. The exactness of the direct transform relies
on the specific choice of weighting functions only depending on $\theta$,
as required by the sampling result on equi-angular grids on the sphere.
Again, the exactness of the evaluation of the inverse scalar spherical
harmonics transform of a band-limited function relies on the fact
that is expressed as the simple finite sum (\ref{eq:sv-03}) with
$l<L$. We will consider the corresponding stable numerical implementations
of the SpharmonicKit package \cite{SAShealy1,SAShealy2}%
\footnote{See www.cs.dartmouth.edu/$\sim$geelong/sphere/ for unpublished documentation
and algorithm implementation (SpharmonicKit package).%
}. As detailed in § \ref{sec:Numerical-implementation}, if high resolutions
are to be considered ($L>1024$), the associated Legendre polynomials
are not pre-calculate in order to avoid a too large RAM memory consumption.
They must be computed on the fly and the corresponding $\mathcal{O}(L^{3})$
time consumption is included in the global asymptotic complexity for
the spherical harmonics transform. Both the $\mathcal{O}(L^{3})$
HEALPix implementation and the $\mathcal{O}(L^{2}\log_{2}^{2}L)$
SpharmonicKit implementation therefore effectively have the same global
$\mathcal{O}(L^{3})$ asymptotic complexity when the calculation of
the associated Legendre polynomials is considered.

Each inverse spin-weighted transform required in (\ref{eq:fs-13})
\emph{a priori} reads as $2^{\vert n\vert}$-terms linear combination
from the recurrence (\ref{eq:fs-23}). But these terms may be grouped
in $\vert n\vert+1$ inverse scalar spherical harmonics transforms
with pre-factors $\cot^{p}\theta/\sin^{q}\theta$ for $p,q\in\mathbb{N}$
and $p+q=\vert n\vert$. In addition, for each $p$ and $q$ with
$p+q=\vert n\vert$, the coefficients for the spins $n$ and $-n$
may be grouped before applying the corresponding inverse scalar spherical
harmonics transform. This allows to consider only spin-weighted transforms
for positive spins and correspondingly further reduces the number
of scalar transforms required for the standard correlation. 

Notice that another recurrence relation might be substituted to (\ref{eq:fs-23})
for the treatment of the inverse spin-weighted transforms through
inverse scalar transforms. This other relation explicitly relates
${}_{n}Y_{lm}$ with ${}_{n\mp1}Y_{lm}$, ${}_{n\mp1}Y_{(l-1)m}$,
and ${}_{n\mp1}Y_{(l+1)m}$ \cite{SASvarshalovich,SASkostelec1}.
The term ${}_{n\mp1}Y_{(l+1)m}(\omega)=[S^{+1}\,\,{}_{n\mp1}Y_{lm}](\omega)$
actually raises the band limit of the associated scalar functions
to be analyzed to $L+\vert n\vert$ after the $\vert n\vert$-steps
recurrence leading from spin-weighted to scalar spherical harmonics,
with $\vert n\vert<L$. From a practical point of view, in a HEALPix
scheme for a fixed resolution, this will reduce the numerical precision
of computation, as the implementation is known to make larger errors
at higher $l$'s. On $2L\times2L$ equi-angular grids, the SpharmonicKit
package is technically limited to consider coefficients up to $L$,
and numerical errors will occur due to the absence of consideration
of the coefficients between $L$ and $L+\vert n\vert$. No such issue
occurs from relation (\ref{eq:fs-23}) which preserves the band limit
$L$ to be considered for the associated scalar functions.

In that context, let us analyze the overall asymptotic complexity
of the standard correlation algorithm proposed. We consider steerable
filters with low azimuthal angular band limit $N<<L$, which constrain
the maximum spin value to $\vert n\vert<N$. Typically $N=2$ for
a first Gaussian derivative, and $N=3$ for a second Gaussian derivative.
The index $n$ therefore decouples from any asymptotic complexity
contribution. The required direct and inverse scalar transforms set
the overall asymptotic complexity of the standard correlation to $\mathcal{O}(L^{3})$
(and even to $\mathcal{O}(L^{2}\log_{2}^{2}L)$ on equi-angular grids
if the associated Legendre polynomials were to be pre-calculated).
The asymptotic complexities of additional contributions to the algorithm
are of order $\mathcal{O}(L^{2})$ and therefore negligible. First,
the pointwise product (\ref{eq:fs-14}) required after the direct
transforms of the signal and the filter is indeed of order $\mathcal{O}(L^{2})$.
Second, the grouping of the $2^{\vert n\vert}$ terms in the recurrence
for each inverse spin-weighted spherical harmonics transform at spin
$n$, as $\vert n\vert+1$ terms on which inverse scalar spherical
harmonics are computed, contribute to the asymptotic complexity as
$2^{\vert n\vert}\times\mathcal{O}(L^{2})$. This contribution is
also negligible for a first or second Gaussian derivative filter.

In summary, our algorithm for the standard correlation of a band-limited
signal with a band-limited steerable filter is structured as follows,
with related asymptotic complexities:

\begin{itemize}
\item Direct scalar spherical harmonics transforms, $\widehat{\Psi}_{ln}$
and $\widehat{F}_{lm}$: $\mathcal{O}(L^{3})$ on all pixelizations
($\mathcal{O}(L^{2}\log_{2}^{2}L)$ on an equi-angular grid if the
associated Legendre polynomials are pre-calculated).
\item Correlation $\widehat{\langle R_{0}\Psi\vert F\rangle}_{lmn}$ in
spherical harmonics space: $\mathcal{O}(L^{2})$ through (\ref{eq:fs-14}).
\item Sum of inverse scalar spherical harmonics transforms to obtain $\langle R_{0}\Psi\vert F\rangle$
through (\ref{eq:fs-13}) and (\ref{eq:fs-23}): $\mathcal{O}(L^{3})$
on all pixelizations ($\mathcal{O}(L^{2}\log_{2}^{2}L)$ on an equi-angular
grid if the associated Legendre polynomials are pre-calculated).
\end{itemize}
Let us finally notice that the reduction of complexity to $\mathcal{O}(L^{2}\log_{2}^{2}L)$
thanks to the use of an equi-angular grid beyond the calculation of
the associated Legendre polynomials provides anyway an asymptotic
reduction of the overall computation times for the corresponding correlation.
It is therefore already of interest only in that respect. But the
explicit calculation of the directional correlation from the relation
(\ref{eq:fs-06}) in terms of the computed standard correlations requires
an additional $\mathcal{O}(L^{3})$ operation. However, we emphasize
that all the information for the directional correlation of a signal
with a steerable filter is already contained in the standard correlations
with its basis filters. In many applications the explicit computation
of the directional correlation through the relation (\ref{eq:fs-06})
is not required. The interest may reside in the computation of the
values of the directional correlation in a small number of specific
directions $\chi_{i}$ at each point $\omega_{0}$ on the sphere.
One may also want to determine the direction $\chi$ at each point
$\omega_{0}$ which maximizes the value of the directional correlation.
In \cite{SASwiaux1}, this last application is illustrated in the
perspective of the detection of the precise direction of local features
in the CMB temperature and polarization anisotropies, through a steerable
wavelet analysis. These features may represent potential signatures
of fundamental non-Gaussianity or statistical anisotropy, or foreground
emissions. The overall computation times for such a directional analysis
are only driven by the asymptotic complexity of the standard correlations.

\subsection{Equi-angular and HEALPix implementations}

\label{sec:Numerical-implementation}

We concisely compare the numerical SpharmonicKit and HEALPix packages
for the scalar spherical harmonics transforms, in terms of memory
requirements, computations times, and numerical stability. Those properties
of the scalar spherical harmonics transforms implementations characterize
the overall standard correlation implementations themselves. We also
comment on the issue of the change of the pixelization on which the
data are sampled, considering equi-angular and HEALPix grids.

Calculations are performed on a $2.20$ GHz Intel Pentium Xeon CPU
with $2$ Gb of RAM memory. Random real test-signals are considered,
with band limits $L\in\{128,256,512,1024\}$. Without loss of generality,
these real test-signals are defined through their scalar spherical
harmonics coefficients $\widehat{G}_{lm}$ with $\vert m\vert\leq l<L$
, with independent real and imaginary parts uniformly distributed
in the interval $[-1,+1]$. The reality condition $\widehat{G}_{l(-m)}=\left(-1\right)^{m}\widehat{G}_{lm}^{*}$
obviously reduces computation times by a factor two relative to generic
complex signals. The inverse and direct scalar spherical harmonics
transforms are successively computed, giving numerical coefficients
$\widehat{H}_{lm}$. The SpharmonicKit implementation is coded in
C. The corresponding inverse transform is computed on $2L\times2L$
equi-angular pixelizations. For $L=1024$, this corresponds to maps
with $N_{pix}=4194304$ pixels. The direct transform is then recomputed
exactly, \emph{i.e.} to the computer's numerical precision, up to
the band limit $L$. The HEALPix implementation considered is coded
in Fortran90. The corresponding inverse transform is computed with
a resolution identified by the parameter $N_{side}=L/2$, identifying
maps with $N_{pix}=12N_{side}^{2}$ equal-area pixels. For $L=1024$,
the value $N_{side}=512$ defines maps with $N_{pix}=3145728$ pixels.
The direct transform is then recomputed with a very good accuracy
up to $L=2N_{side}$.

RAM memory requirements for the scalar transforms are as follows.
If the required associated Legendre polynomials $P_{l}^{m}(\cos\theta)$
are pre-calculated once for all values of $l$, $\theta$, and $m$,
and stored in RAM memory, the number of real values of associated
Legendre polynomials $P_{l}^{m}(\cos\theta)$ stored in RAM memory
for all $l$, $\theta$, and $m$ is of order $\mathcal{O}(L^{3})$.
The overall memory requirements allowing both direct and inverse transforms
are still easily accessible but already higher than $1$ Gb for $L=1024$,
for double-precision numbers. The pre-calculation computation time
itself is of order $\mathcal{O}(L^{3})$ through the use of a recurrence
relation in $l$ on the associated Legendre polynomials. As a pre-calculation,
the corresponding time consumption is not to be taken into account
in the reported computation times, which consequently remain of order
$\mathcal{O}(L^{2}\log_{2}^{2}L)$ and $\mathcal{O}(L^{3})$ in the
SpharmonicKit and the HEALPix implementations respectively. However,
the corresponding memory requirements rapidly become unaffordable
for resolutions higher than $L=1024$, such as the resolution expected
for the Planck experiment. We therefore chose the SpharmonicKit and
HEALPix implementations for which the required associated Legendre
polynomials $P_{l}^{m}(\cos\theta)$ are calculated on the fly, and
stored in RAM memory for all values of $l$ and $\theta$, but for
only one value of $m$ at the time. The reported computation times
therefore inevitably include the $\mathcal{O}(L^{3})$ calculation
of the Legendre polynomials, to be added to the remaining $\mathcal{O}(L^{2}\log_{2}^{2}L)$
and $\mathcal{O}(L^{3})$ terms in the SpharmonicKit and the HEALPix
implementations respectively. Both implementations effectively acquire
the same global $\mathcal{O}(L^{3})$ asymptotic complexity. For fixed
$m$, the number of real values $P_{l}^{m}(\cos\theta)$ stored in
RAM memory for all values of $l$ and $\theta$ is $\mathcal{O}(L^{2})$.
For the SpharmonicKit and HEALPix implementations, the overall memory
requirements allowing both direct and inverse transforms are respectively
of order $1.3\times10^{2}$ Mb and $4.6\times10^{1}$ Mb for $L=1024$,
for double-precision numbers. They are therefore significantly lower
than the corresponding values when the associated Legendre polynomials
are pre-calculated, and Planck-type resolutions are easily affordable.

The related computation times associated with these global $\mathcal{O}(L^{3})$
SpharmonicKit and HEALPix implementations for the inverse and direct
transforms are given in Table \ref{cap:scalartransform-times}. They
are averages over $5$ random band-limited real test-signals. They
are of the order of tens of seconds at $L=1024$ both for the SpharmonicKit
and the HEALPix implementations, in rough agreement with our previous
intuitive estimation that an $\mathcal{O}(L^{2})$ operation requires
$0.03$ seconds on our standard computer. For the SpharmonicKit implementation,
computation times evolve from $1.0\times10^{-1}$ seconds for $L=128$
to $4.6\times10^{1}$ seconds for $L=1024$. For the HEALPix implementation,
notice that an iterative scheme can be used for the direct transform,
in order to enhance the accuracy of the calculation. After several
iterations, the process converges towards a limit precision, which
however never reaches the precision of the SpharmonicKit approach.
For zero iteration ($i=0$), computation times evolve from $4.8\times10^{-2}$
seconds for $L=128$ to $1.5\times10^{1}$ seconds for $L=1024$.
The HEALPix implementation with a unique iteration is therefore slightly
more rapid than the SpharmonicKit one, at least up to the band limit
$L=1024$. Computation times grow significantly with the number $i$
of iterations used. For one iteration ($i=1$), the computation times
for the direct transform are already bigger than for the SpharmonicKit
implementation. They evolve from $2.7\times10^{-1}$ seconds for $L=128$
to $7.8\times10^{1}$ seconds for $L=1024$.

\begin{table}

\begin{center}

\caption{Scalar transforms computation times \label{cap:scalartransform-times}}

\begin{tabular}{ccccc}

\hline\hline

&Time $L=128$&Time $L=256$&Time $L=512$&Time $L=1024$\\

&(sec)&(sec)&(sec)&(sec)\\

\hline

Spharmonic-&$1.0\textnormal{e}-01$&$6.8\textnormal{e}-01$&$5.2\textnormal{e}+00$&$4.2\textnormal{e}+01$\\

Kit&$9.5\textnormal{e}-02$&$6.7\textnormal{e}-01$&$5.5\textnormal{e}+00$&$4.6\textnormal{e}+01$\\

\hline

HEALPix2.1&$4.8\textnormal{e}-02$&$3.0\textnormal{e}-01$&$2.0\textnormal{e}+00$&$1.5\textnormal{e}+01$\\

($i=0$)&$4.0\textnormal{e}-02$&$2.5\textnormal{e}-01$&$1.6\textnormal{e}+00$&$1.1\textnormal{e}+01$\\

\hline

HEALPix2.1&$2.7\textnormal{e}-01$&$1.7\textnormal{e}+00$&$1.1\textnormal{e}+01$&$7.8\textnormal{e}+01$\\

($i=1$)&--&--&--&--\\

\hline

\end{tabular}

\end{center}

\tablecomments{Computation times for the SpharmonicKit and HEALPix2.1 implementations of the scalar spherical harmonics transforms, measured on a $2.20$ GHz Intel Pentium Xeon CPU with $2$ Gb of RAM memory. Times associated with the direct transform are listed above corresponding times for the inverse transform. The HEALPix2.1 direct transforms are reported both at $i=0$ and $i=1$ iteration.}

\end{table}

The relative maximum error on the coefficients associated with the
scalar spherical harmonics transforms is defined as $\max_{l,m}\vert(\widehat{G}_{lm}-\widehat{H}_{lm})/\widehat{G}_{lm}\vert$,
where $\vert\cdot\vert$ here denotes the complex norm. The relative
root mean square numerical error is defined as the ratio of the $L^{2}$-norm
of $G-H$ and $G$, that is through the Plancherel relation: $(\sum_{l,m}\vert\widehat{G}_{lm}-\widehat{H}_{lm}\vert^{2}/\sum_{l,m}\vert\widehat{G}_{lm}\vert^{2})^{1/2}$.
The numerical errors given in Table \ref{cap:scalartransform-errors}
are averages for transforms over $5$ random band-limited real test-signals.
For the SpharmonicKit implementation, errors are extremely small as
expected from the theoretical exactness of the algorithm. Relative
maximum errors on the coefficients and relative root mean square errors
are indeed bounded by $1.8\times10^{-7}$ and $7.5\times10^{-10}$
respectively, up to $L=1024$. For the HEALPix implementation with
zero iteration ($i=0$), the relative maximum errors on the coefficients
can reach $2.9\times10^{1}$, up to $L=1024$. But the relative root
mean square errors are bounded by $7.0\times10^{-3}$, up to $L=1024$,
which ensures the global numerical precision of the implementation.
This already illustrates the numerical stability of the HEALPix implementation
at the $0.7$ \% level, which however remains a much lower precision
than with the SpharmonicKit implementation. The use of the iterative
scheme for the computation of the direct transform allows to reach
a better accuracy, at the price of a bigger time consumption. For
the HEALPix implementation with one iteration ($i=1$), the relative
maximum errors on the coefficients and relative root mean square errors
are then bounded by $1.1$ and $2.7\times10^{-4}$ respectively, up
to $L=1024$, therefore achieving a numerical stability at the $0.027$
\% level.

\begin{table}

\begin{center}

\caption{Scalar transforms error analysis \label{cap:scalartransform-errors}}

\begin{tabular}{ccccc}

\hline\hline

&Error $L=128$&Error $L=256$&Error $L=512$&Error $L=1024$\\

\hline

Spharmonic-&$3.3\textnormal{e}-11$&$9.4\textnormal{e}-11$&$2.6\textnormal{e}-10$&$7.5\textnormal{e}-10$\\

Kit&$2.9\textnormal{e}-09$&$3.7\textnormal{e}-09$&$1.1\textnormal{e}-08$&$1.8\textnormal{e}-07$\\

\hline

HEALPix2.1&$7.0\textnormal{e}-03$&$4.8\textnormal{e}-03$&$3.5\textnormal{e}-03$&$2.4\textnormal{e}-03$\\

($i=0$)&$7.7\textnormal{e}+00$&$1.0\textnormal{e}+01$&$2.9\textnormal{e}+01$&$2.3\textnormal{e}+01$\\

\hline

HEALPix2.1&$2.7\textnormal{e}-04$&$1.8\textnormal{e}-04$&$1.4\textnormal{e}-04$&$9.2\textnormal{e}-05$\\

($i=1$)&$2.7\textnormal{e}-01$&$3.6\textnormal{e}-01$&$1.1\textnormal{e}+00$&$7.7\textnormal{e}-01$\\

\hline

\end{tabular}

\end{center}

\tablecomments{Errors for the SpharmonicKit and HEALPix2.1 implementations of the scalar spherical harmonics transforms, measured on a $2.20$ GHz Intel Pentium Xeon CPU with $2$ Gb of RAM memory. Relative root mean square errors after inverse and direct transforms are listed above the corresponding relative maximum errors on the coefficients. HEALPix2.1 results are reported both at $i=0$ and $i=1$ iteration.}

\end{table}

Let us emphasize that the conclusions of this comparison of memory
requirements, computation times, and numerical stability remain for
the equi-angular and HEALPix implementations of our standard correlation
algorithm. Indeed, the overall computation is strictly driven by the
SpharmonicKit and HEALPix implementations of the scalar spherical
harmonics transforms respectively.

We finally briefly comment on the issue of changing the pixelization
on which the signal is defined, from HEALPix to equi-angular, or conversely.
A specific choice of pixelization might be preferred for some analyses.
The above results confirm the very good precision with which both
implementations, with resolution $N_{side}=L/2$ for HEALPix and on
$2L\times2L$ equi-angular grids for SpharmonicKit, can calculate
the direct and inverse transforms of a signal up to the band limit
$L$. Consequently, a change of pixelization may be performed safely
by first computing the spherical harmonics coefficients of the signal
by direct transform from one pixelization, and second operate an inverse
transform on the alternative pixelization. The numerical precision
of such an operation is obviously limited by the transform with the
lowest precision, \emph{i.e.} the HEALPix transform.

Notice however that any spectral or directional correlation analysis
only uses the spherical harmonics coefficients of the signal. In that
perspective, the question of pixelization change makes no sense. The
initial pixelization on which the signal is defined does not constrain
such analyses as both the HEALPix and SpharmonicKit implementations
allow a precise calculation of the spherical harmonics coefficients
of the signal. But the better numerical precision and lower computation
times of the SpharmonicKit transform can really materialize if the
data are originally sampled on equi-angular grids. Such a possibility
must be evaluated for each application independently. The question
can notably be raised for CMB experiments such as WMAP and Planck,
which currently use HEALPix grids \cite{SASgorski}.

\subsection{Existing alternative procedure and optimization with steerable filters}

For completeness of the discussion, we briefly expose an existing
alternative algorithm for the directional correlation, based on the
factorization of the rotation operators $R(\rho)$ on functions in
$L^{2}(S^{2},d\Omega)$ on the sphere as \cite{SASrisbo,SASwandelt}\begin{equation}
R\left(\varphi_{0},\theta_{0},\chi\right)=R\left(\varphi_{0}-\frac{\pi}{2},-\frac{\pi}{2},\theta_{0}\right)R\left(0,\frac{\pi}{2},\chi+\frac{\pi}{2}\right).\label{eq:sv-12}\end{equation}
 The directional correlation relation (\ref{eq:sv-10}) and the expression
(\ref{eq:sv-05}) of the Wigner $D$-functions, matrix elements of
the operators $R(\rho)$, therefore give an alternative expression
for the directional correlation of arbitrary signals $F$ and (non-steerable)
filters $\Psi$ on the sphere. We get indeed\begin{equation}
\langle R\left(\rho\right)\Psi\vert F\rangle=\sum_{m,m',n\in\mathbb{Z}}\widehat{\langle R\Psi\vert F\rangle}_{mm'n}e^{i(m\varphi_{0}+m'\theta_{0}+n\chi)},\label{eq:sv-13}\end{equation}
with the Fourier coefficients on the three-torus $\widehat{\langle R\Psi\vert F\rangle}_{mm'n}$
given by\begin{equation}
\widehat{\langle R\Psi\vert F\rangle}_{mm'n}=e^{i(n-m)\pi/2}\sum_{l\geq C}d_{m'm}^{l}\left(\frac{\pi}{2}\right)d_{m'n}^{l}\left(\frac{\pi}{2}\right)\widehat{\Psi}_{ln}^{*}\widehat{F}_{lm},\label{eq:sv-14}\end{equation}
where $C=\max(\vert m\vert,\vert m'\vert,\vert n\vert)$, and with
the symmetry relation $d_{m'm}^{l}(\theta)=d_{mm'}^{l}(-\theta)$
\cite{SASvarshalovich}. For a band-limited signal $F$ with band
limit $L$ on the sphere one has $\vert m\vert,\vert m'\vert,\vert n\vert\leq l<L$.
In these terms, the directional correlation algorithm implemented
through the factorization of rotations is structured as follows, with
related asymptotic complexities:

\begin{itemize}
\item Direct scalar spherical harmonics transforms, $\widehat{\Psi}_{ln}$
and $\widehat{F}_{lm}$: $\mathcal{O}(L^{3})$ on all pixelizations
($\mathcal{O}(L^{2}\log_{2}^{2}L)$ on an equi-angular grid if the
associated Legendre polynomials are pre-calculated).
\item Correlation $\widehat{\langle R\Psi\vert F\rangle}_{mm'n}$ in spherical
harmonics space: $\mathcal{O}(L^{4})$ through (\ref{eq:sv-14}).
\item Inverse Fourier transform $\langle R\Psi\vert F\rangle$ on the three-torus
in $(\theta_{0},\varphi_{0})$: $\mathcal{O}(L^{3}\log_{2}L)$ on
an equi-angular grid by the use of the standard Cooley-Tukey fast
Fourier transform algorithm, while $\mathcal{O}(L^{4})$ on HEALPix
and other pixelizations.
\end{itemize}
The overall asymptotic complexity is therefore of order $\mathcal{O}(L^{4})$
independently of the pixelization, due to the correlation in spherical
harmonics space.

Let us reconsider our first naive estimation of the computation times
for the directional correlation on maps of several megapixels on the
sphere. For $L\simeq10^{3}$, an $\mathcal{O}(L^{4})$ algorithm already
greatly reduces computation times from years down to days on a single
standard computer. However, if a large number of signals have to be
considered, say thousands of simulations, this remains hardly affordable
even through the use of multiple computers.

Our algorithm achieves an $\mathcal{O}(L^{3})$ asymptotic complexity
for the standard or directional correlation with steerable filters,
and even an $\mathcal{O}(L^{2}\log_{2}^{2}L)$ asymptotic complexity
on equi-angular grids if the associated Legendre polynomials are pre-calculated.
The algorithm described here using the factorization of rotations
can also achieve an $\mathcal{O}(L^{3})$ complexity for the standard
or directional correlation, if steerable filters are considered. Indeed,
for a steerable filter $\Psi$ with an azimuthal band limit $N$,
the index $n$ decouples from asymptotic complexity counts: $\vert n\vert<N<<L$
. But the corresponding structure does not allow any reduction to
$\mathcal{O}(L^{2}\log_{2}^{2}L)$, as it clearly appears from relation
(\ref{eq:sv-14}).

Notice that the first proposal for the directional wavelet analysis
of the CMB \cite{WNGmcewen1,WNGmcewen2} was based on the presently
discussed algorithm using the factorization of rotations. Instead
of considering steerable wavelet filters, a much less precise sampling
in $P$ points in the direction $\chi$ was defined, with $P<<L$,
typically reducing the precision in the directional analysis from
$\Delta\chi=2\pi/L$ to $\Delta\chi=2\pi/P$. This also artificially
applies a band limit in the related Fourier index $n$: $\vert n\vert<P$.
Technically, the asymptotic complexity of directional correlation
is thus reduced like if a steerable filter with an azimuthal band
limit $P$ was considered. However, this is by no means an answer
to the problem of the directional correlation of band-limited signals
on the sphere if one requires the same high precision ($P\simeq L$)
in the analysis of local directions as for the position of features
on the sphere. In that perspective, the use of steerable filters is
essential. As discussed in \cite{SASwiaux1}, the theoretical angular
resolution power of a steerable filter is infinite in the sense that,
around each point of the signal, it can resolve exactly the direction
of local features with a pre-defined morphology. It therefore makes
complete sense to define a precise sampling in the analysis of local
directions ($P\simeq L$). If needed, the explicit calculation of
the directional correlation for all sampled values of $\chi$ therefore
requires (see relation (\ref{eq:fs-06})) the additional $\mathcal{O}(P\times L^{2})\simeq\mathcal{O}(L^{3})$
operation quoted in subsection \ref{sub:Detailed-algorithmic-structure}.
Obviously, the resolution of the angular morphology itself of a local
feature is limited by the non-zero angular width of the steerable
filter, \emph{i.e.} its angular band limit $N$ in the index $n$
associated with the azimuthal angle $\varphi$ when the filter is
located at the North pole.

\section{Application to the CMB Scale-space analysis}

\label{sec:CMB-Scale-space-analysis}

\subsection{Temperature and polarization}

The CMB radiation is completely described in terms of the observable
temperature $T$ and linear polarization $Q$ and $U$ Stokes parameters
\cite{CMBPOLseljak,CMBzaldarriaga,CMBPOLkosowsky,CMBPOLkamionkowski1,CMBPOLkamionkowski2,CMBPOLhu}.
On the one hand, the temperature is a scalar function on the sphere,
\emph{i.e.} invariant under local rotations in the plane tangent to
the sphere at each point. It is also invariant under global inversion
of the three-dimensional coordinates. On the other hand, the $Q$
and $U$ Stokes parameters are not invariant, but transform as the
components of a transverse, symmetric, and traceless rank $2$ tensor
on the sphere, and respectively have even and odd parities under global
inversion. Stated in other words, the combinations $Q\pm iU$ are
spin $\pm2$ functions on the sphere, and transform in one another
under global inversion. In this context, the directional correlation
of the temperature field with a scalar filter is invariant under the
considered coordinates transformations. The directional correlations
of the linear polarization parameters $Q$ and $U$ with a scalar
filter are obviously not invariant, but transform in one another just
as $Q$ and $U$. A physical analysis will require the definition
of invariant quantities. In that respect, the linear polarization
of the CMB is equivalently defined by its electric $E$ and magnetic
$B$ components, uniquely defined in terms of $Q$ and $U$. These
two polarization components are scalar functions on the sphere and
also respectively have even and odd parities under global inversion.
In this context, the statistics of a Gaussian and stationary CMB signal
is completely described by the four invariant temperature ($TT$),
polarization ($EE$ and $BB$), and cross-correlation ($TE$) angular
power spectra. In the same idea, the complete scale-space analysis
of the CMB data through directional correlation on the sphere will
therefore rely the directional correlation of the scalar temperature
$T$, and the scalar polarization components $E$ and $B$, with an
arbitrary scalar filter.

From the numerical point of view, $E$ and $B$ are related to the
observables $Q$ and $U$ through derivative relations. It is consequently
not possible to transform the experimental $Q$ and $U$ maps in scalar
$E$ and $B$ maps from their basic definitions. However, the scalar
spherical harmonics coefficients $\widehat{E}_{lm}$ and $\widehat{B}_{lm}$
are explicitly given as simple linear combinations of the spin $\pm2$
spherical harmonics coefficients ${}_{\pm2}\widehat{(Q\pm iU)}_{lm}$.
From the recurrence relation (\ref{eq:fs-23}), these spin $n=\pm2$
spherical harmonics coefficients may computed as linear combinations
of $\vert n\vert+1=3$ scalar spherical harmonics coefficients. The
direct scalar spherical harmonics transforms of $E$ and $B$ can
therefore be obtained from the experimental $Q$ and $U$ maps with
the same asymptotic complexity as the direct scalar spherical harmonics
transform of an experimental $T$ map. From that stage, the remaining
part of the detailed algorithmic structure proposed in § \ref{sub:Detailed-algorithmic-structure}
can be applied identically to $\widehat{E}_{lm}$, $\widehat{B}_{lm}$,
or $\widehat{T}_{lm}$. In summary, our algorithm can be applied both
for the directional correlation of temperature and polarization CMB
maps with steerable filters. The overall asymptotic complexity for
the analysis of polarization maps remains $\mathcal{O}(L^{3})$ on
all pixelizations ($\mathcal{O}(L^{2}\log_{2}^{2}L)$ on an equi-angular
grid if the associated Legendre polynomials are pre-calculated).

\subsection{Numerical illustration}

As an illustration we apply our algorithm to the computation of the
wavelet coefficients of a temperature map with a second Gaussian derivative
wavelet. A CMB temperature map is simulated both in the equi-angular
and HEALPix pixelizations, from the angular power spectrum which best
fits the three-year WMAP data. A comparison of the implementations
of our algorithm on the two grids is proposed both in terms of computation
times, and precision of analysis.

First, we produce a simulated CMB temperature map both on equi-angular
and HEALPix pixelizations in the following way. We define scalar spherical
harmonics coefficients $\widehat{T}_{lm}$ from a Gaussian distribution
with zero mean and a variance given by the temperature angular power
spectrum $C_{l}^{TT}$ which best fits the three-year WMAP data \cite{CMBspergel2},
up to a band limit $L=1024$. The spectrum $C_{l}^{TT}$ was computed
with CMBFAST%
\footnote{See cmbfast.org for documentation (CMBFAST4.5.1 software).%
}. Two real CMB temperature maps are then produced by inverse scalar
spherical harmonics transforms of these coefficients, on a $2L\times2L$
equi-angular grid with the SpharmonicKit, and with a resolution $N_{side}=L/2$
for the HEALPix pixelization.

Second, we apply our algorithm to produce the directional correlation
of each map with a second Gaussian derivative wavelet. The equi-angular
or HEALPix implementations of our algorithm are used independently
on the two corresponding maps. For coherence with the studied scalar
spherical harmonics transforms codes, the complete equi-angular implementation
is coded in C, while the HEALPix implementation is coded in Fortran90.

For a given scale $a\in\mathbb{R}_{+}^{*}$, the angular size of our
dilated wavelets $\Psi_{a}$ on the sphere is defined as twice the
dispersion of the corresponding Gaussian. Fifteen scales are selected
corresponding to angular sizes of the second Gaussian derivative lying
between $20$ arcminutes and $60$ degrees: $a_{1}\rightarrow20'$,
$a_{2}\rightarrow40'$, $a_{3}\rightarrow1^{\circ}$, $a_{4}\rightarrow5^{\circ}$,
$a_{5}\rightarrow10^{\circ}$, $a_{6}\rightarrow15^{\circ}$, $a_{7}\rightarrow20^{\circ}$,
$a_{8}\rightarrow25^{\circ}$, $a_{9}\rightarrow30^{\circ}$ $a_{10}\rightarrow35^{\circ}$,
$a_{11}\rightarrow40^{\circ}$, $a_{12}\rightarrow45^{\circ}$, $a_{13}\rightarrow50^{\circ}$,
$a_{14}\rightarrow55^{\circ}$, $a_{15}\rightarrow60^{\circ}$. The
analytical expression of the second Gaussian derivative wavelet as
a function of the scale $a$ may be found in \cite{SASwiaux1}. From
a practical point of view, the wavelet is first sampled on the sphere
in each pixelization, and the corresponding spherical harmonics coefficients
$(\widehat{\Psi_{a}})_{lm}$ are calculated by direct spherical harmonics
transform at each scale.

The wavelet coefficients resulting from the directional correlation
of the signal with the second Gaussian derivative at each scale $a$
read as $W_{\Psi}^{T}(\chi,\omega_{0},a)=\langle R\left(\rho\right)\Psi_{a}\vert T\rangle$,
for $\rho=(\varphi_{0},\theta_{0},\chi)$, and $\omega_{0}=(\varphi_{0},\theta_{0})$.
The coefficients resulting from the standard correlations with the
corresponding basis filters defined in (\ref{eq:fs-05}) read $W_{\Psi_{m}}^{T}(\omega_{0},a)=\langle R(\omega_{0})\Psi_{m\, a}\vert T\rangle$,
for $1\leq m\leq3$, and with $\Psi_{1\, a}=\Psi_{a}^{\partial_{\hat{x}}^{2}(gauss)}$,
$\Psi_{2\, a}=\Psi_{a}^{\partial_{\hat{y}}^{2}(gauss)}$, and $\Psi_{3\, a}=\Psi_{a}^{\partial_{\hat{x}}\partial_{\hat{y}}(gauss)}$.
The steerability relation (\ref{eq:fs-06}) therefore explicitly reads
\begin{equation}
W_{\Psi}^{T}\left(\chi,\omega_{0},a\right)=\sum_{m=1}^{3}k_{m}\left(\chi\right)W_{\Psi_{m}}^{T}\left(\omega_{0},a\right),\label{eq:ni-01}\end{equation}
with the weights $k_{1}(\chi)=\cos^{2}\chi$, $k_{2}(\chi)=\sin^{2}\chi$,
and $k_{3}(\chi)=\sin2\chi$ given in (\ref{eq:fs-04}). At each of
the fifteen scales considered, the coefficients $W_{\Psi_{m}}^{T}(\omega_{0},a)$
gather all the information necessary to compute the directional correlation
of the signal with the second Gaussian derivative. These coefficients
are computed as functions on the sphere, up to the band limit $L=1024$.

On the one hand, the overall computation times at each scale for the
coefficients $W_{\Psi_{m}}^{T}(\omega_{0},a)$ of the three basis
filters ($1\leq m\leq3$) are of the order of $7.1\times10^{2}$ seconds
in the equi-angular implementation on a $2.20$ GHz Intel Pentium
Xeon CPU with $2$ Gb of RAM memory. The same calculation takes $2.1\times10^{2}$
seconds in the HEALPix implementation wih zero iteration ($i=0$)
for the direct transforms. These values may directly be inferred from
the computation times reported in Table \ref{cap:scalartransform-times}
for the scalar spherical harmonics transforms. Indeed, the second
Gaussian derivative contains the frequencies $n=\{0,\pm2\}$ and $\vert n\vert+1$
inverse scalar spherical harmonics are required for each spin-weighted
transform in (\ref{eq:fs-13}). Let us also recall that opposite spins
are grouped by the algorithm before applying the scalar spherical
harmonics transforms. For each of the three basis filters, four inverse
scalar transforms (one at spin $0$ and three at spin $2$) therefore
add up to the direct scalar transform of the filter. Also adding the
direct scalar transform of the signal, a total of four direct scalar
transforms and twelve inverse scalar transforms are required, which
leads to a very good estimation of the overall computation times reported
above. 

On the other hand, the relative precision of the equi-angular and
HEALPix implementations may be probed quantitatively through the calculation
of the mean and variance of the wavelet coefficients at each scale
$a$ and each direction $\chi$, respectively\begin{eqnarray}
\mu_{\Psi}^{T}\left(\chi,a\right) & = & \frac{1}{4\pi}\int_{S^{2}}d\Omega_{0}W_{\Psi}^{T}\left(\chi,\omega_{0},a\right)\nonumber \\
\left[\sigma_{\Psi}^{T}\right]^{2}\left(\chi,a\right) & = & \frac{1}{4\pi}\int_{S^{2}}d\Omega_{0}\vert W_{\Psi}^{T}\left(\chi,\omega_{0},a\right)-\mu_{\Psi}^{T}\left(\chi,a\right)\vert^{2}.\label{eq:ni-02}\end{eqnarray}
 From relation (\ref{eq:ni-01}), those moments are explicitly given
from the means and (co-)variances of the standard correlations at
each scale $a$, respectively, as\begin{eqnarray}
\mu_{\Psi}^{T}\left(\chi,a\right) & = & \sum_{m=1}^{3}k_{m}\left(\chi\right)\mu_{m}^{T}\left(a\right)\nonumber \\
\left[\sigma_{\Psi}^{T}\right]^{2}\left(\chi,a\right) & = & \sum_{m,m'=1}^{3}k_{m}^{*}\left(\chi\right)k_{m'}\left(\chi\right)\sigma_{mm'}^{T}\left(a\right),\label{eq:ni-03}\end{eqnarray}
with\begin{eqnarray}
\mu_{m}^{T}\left(a\right) & = & \frac{1}{4\pi}\int_{S^{2}}d\Omega_{0}W_{\Psi_{m}}^{T}\left(\omega_{0},a\right)\nonumber \\
\sigma_{mm'}^{T}\left(a\right) & = & \frac{1}{4\pi}\int_{S^{2}}d\Omega_{0}\left(W_{\Psi_{m}}^{T}\left(\omega_{0},a\right)-\mu_{\Psi_{m}}^{T}\left(a\right)\right)^{*}\nonumber \\
 &  & \left(W_{\Psi_{m'}}^{T}\left(\omega_{0},a\right)-\mu_{\Psi_{m'}}^{T}\left(a\right)\right).\label{eq:ni-04}\end{eqnarray}
All these quantities are defined from the directional correlation
of a real signal with a real wavelet, and are therefore themselves
real. Notice that quadrature rules need to be applied in order to
compute the latter integrals on both pixelizations. On equi-angular
grids, the sampling theorem discussed in §\ref{sec:Fast-algorithm}
equivalently states that the integral of a scalar function on the
sphere with a band limit $2L$ may be computed exactly as a weighted
sum of its sampled values on $2L\times2L$ equi-angular grids. The
quadrature weights are the same as those applied for the computation
of spherical harmonics coefficients \cite{SASdriscoll}. On HEALPix
grids, the integral of a function on the sphere is simply approximated
by the sum of its sampled values, as all pixels have the same area.
These rules are applied for the computation of the integrals $\mu_{m}^{T}(a)$
and $\sigma_{mm'}^{T}(a)$ on the sphere, at each of the fifteen scales
considered.

\begin{figure}

\begin{center}

\includegraphics[width=8.5cm]{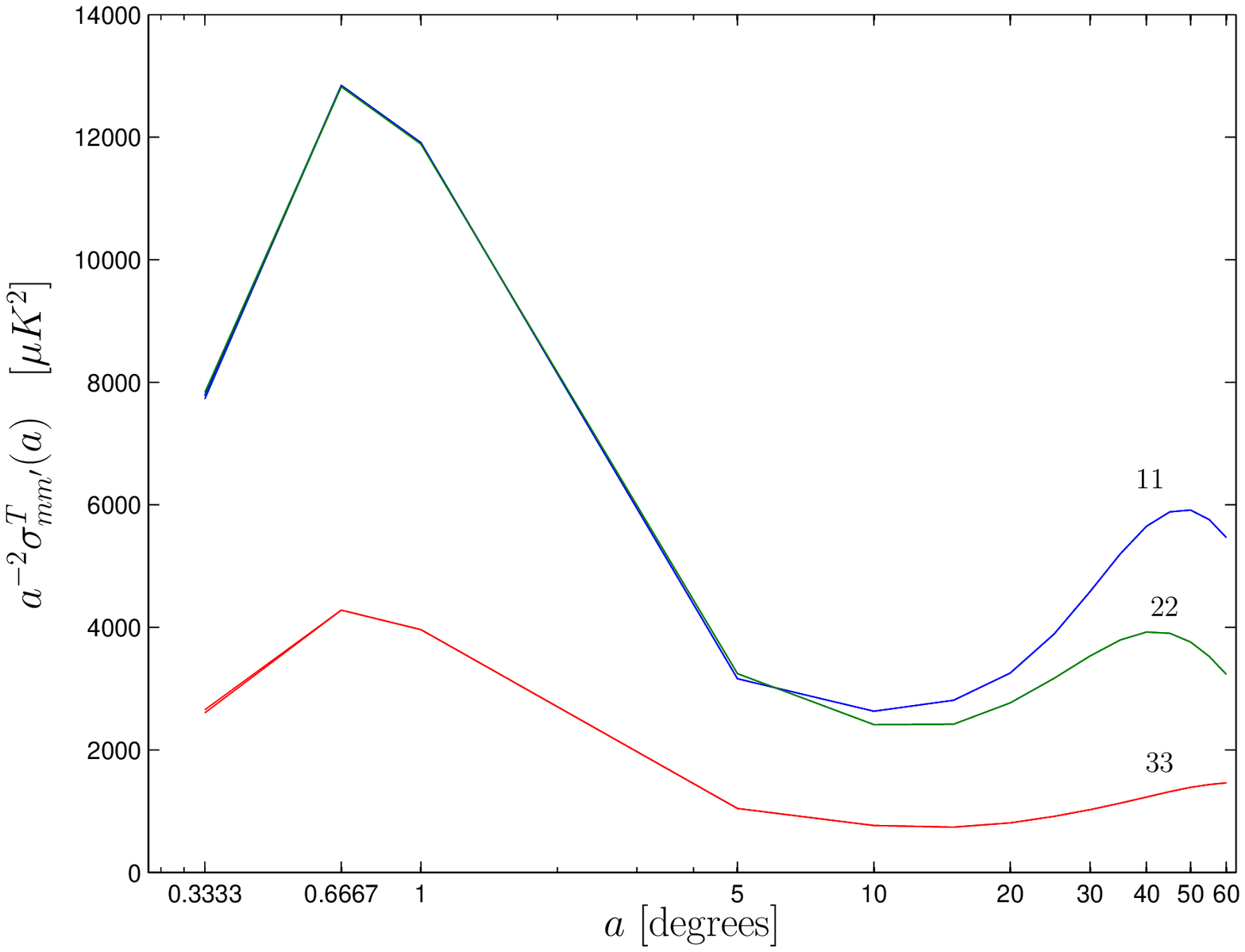}\\

\includegraphics[width=8.5cm]{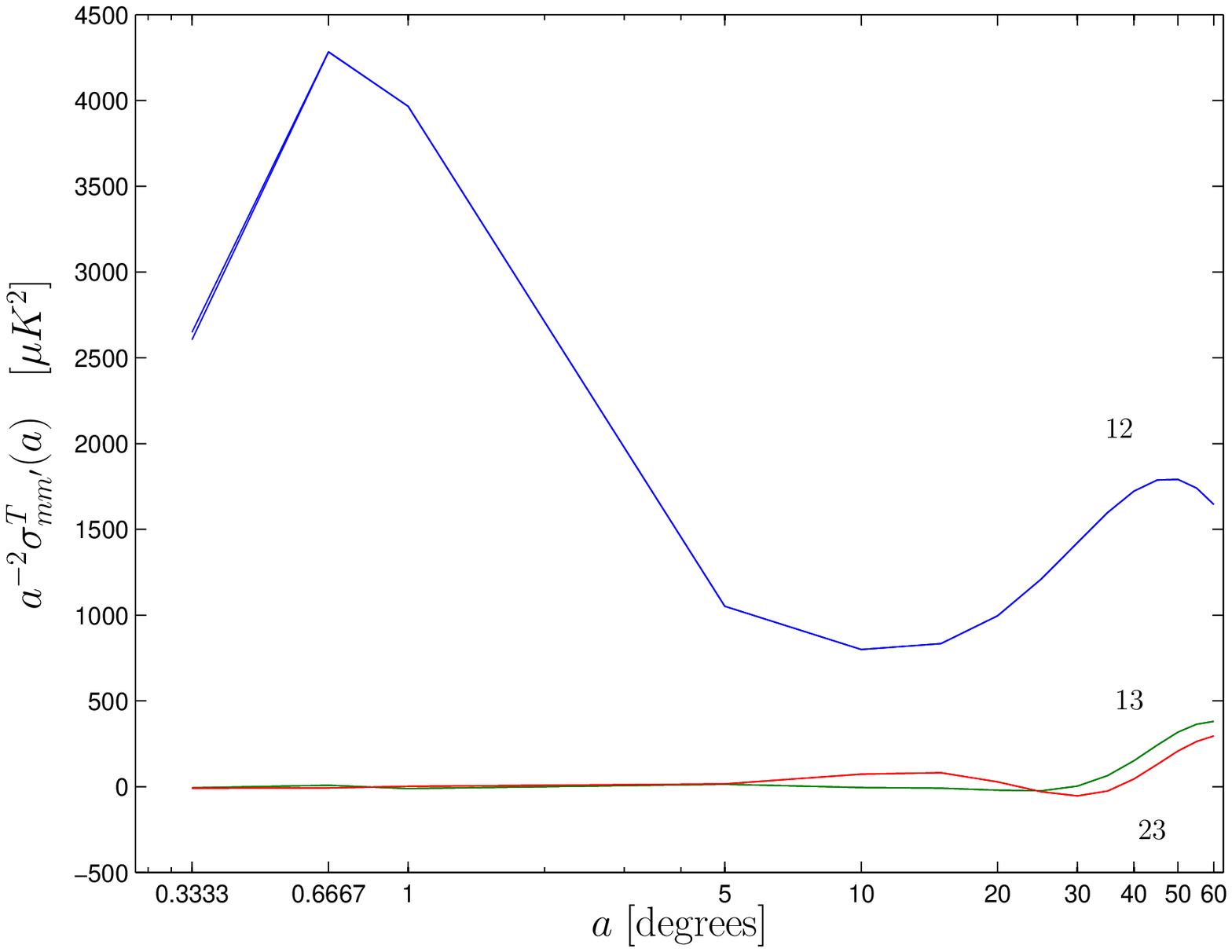}

\caption{\label{cap:comparison-variances-L2log2L-L3} Comparison of the relative precision of the equi-angular and HEALPix implementations of the fast directional correlation algorithm. The six spectra associated with the wavelet coefficients from the directional correlation of a simulated CMB temperature map with the second Gaussian derivative wavelet are shown. The spectra, normalized as $a^{-2}\sigma^T_{mm'}(a)$ with $1 \leq m,m' \leq 3$, are given in $\mu K^2$ as functions of the wavelet scale $a$, converted in terms of angular size in degrees. The top panel represents the variances $\sigma^T_{11}(a)$, $\sigma^T_{22}(a)$, and $\sigma^T_{33}(a)$. The bottom panel represents the covariances $\sigma^T_{12}(a)$, $\sigma^T_{13}(a)$, and $\sigma^T_{23}(a)$. For all six spectra and at any of the fifteen scales considered, the maximum relative difference between the two implementations is bounded by $1.9$ \%. The equi-angular and HEALPix curves are therefore indistinguishable.}

\end{center}

\end{figure}

Figure \ref{cap:comparison-variances-L2log2L-L3} represents the six
spectra $\sigma_{mm'}^{T}(a)$ computed in the equi-angular and HEALPix
implementations, as functions of the scale. The top panel represents
the three variances $\sigma_{mm}^{T}(a)$, with $1\leq m\leq3$. The
bottom panel represents the three covariances $\sigma_{mm'}^{T}(a)$
with $1\leq m\neq m'\leq3$.For any of those six spectra, at the first scale $a_{1}\rightarrow20'$,
the maximum relative difference between the values arising from the
equi-angular and HEALPix implementations is bounded by $1.9$ \%.
At that small scale, the assumption that the wavelet filter is band-limited
at $L$ is not perfect anymore. The wavelet is not perfectly well
sampled on the grids considered. Consequently, aliasing errors can
occur in both the equi-angular and HEALPix implementations, which
explain the discrepancy between the spectra. For the remaining fourteen
scales, the maximum relative difference between the spectra from the
equi-angular and HEALPix implementations drops below $0.07$ \%. This
remaining difference may be attributed to the calculation of the scalar
spherical harmonics transforms in the HEALPix implementation, which
is approximate even for band-limited signals and filters. This discrepancy
is indeed coherent with the relative root mean square error on the
spherical harmonics transform, estimated at $0.24$ \% in Table \ref{cap:scalartransform-errors}
for HEALPix with zero iteration ($i=0$), and at $L=1024$. We also
know that this already very good precision of the HEALPix implementation
can be enhanced using at least one iteration ($i\geq1$), at the price
of an increased computation time. This would reduce the relative differences
with the theoretically exact equi-angular implementation.

\section{Conclusion}

\label{sec:Conclusion}We introduced a fast algorithm for the directional
correlation of band-limited signals with band-limited steerable filters
on the sphere. The \emph{a priori} asymptotic complexity associated
with the directional correlation is $\mathcal{O}(L^{5})$, where $2L$
stands for the square-root of the number of sampling points on the
sphere, also setting a band limit $L$ for the signals and filters
considered. The use of steerable filters allows to compute the directional
correlation uniquely in terms of direct and inverse scalar spherical
harmonics transforms. The overall asymptotic complexity is correspondingly
reduced to the asymptotic complexity of scalar spherical harmonics
transforms, that is $\mathcal{O}(L^{3})$. We also emphasized that
an $\mathcal{O}(L^{2}\log_{2}^{2}L)$ asymptotic complexity may be
achieved on equi-angular pixelizations if associated Legendre polynomials
are pre-calculated. We thoroughly compared the implementations of
the scalar spherical harmonics transforms on HEALPix and equi-angular
grids, in terms of memory requirements, numerical stability, and computation
times. First, the memory requirements are easily accessible. Second,
the SpharmonicKit implementation on equi-angular grids based on an
exact algorithm proposed by Driscoll and Healy reaches the computer's
numerical precision. The exactness of the calculation relies on a
sampling theorem for scalar functions on equi-angular grids on the
sphere. The already good precision of the HEALPix implementation of
the scalar spherical harmonics transform may be enhanced through an
iterative process, but never reaches the level achieved on equi-angular
grids. Third, the computation times for the $\mathcal{O}(L^{3})$
scalar transform of maps of several megapixels on the sphere ($L\simeq10^{3}$)
are reduced from years to tens of seconds on a single standard computer
in both the equi-angular and HEALPix implementations. The equi-angular
implementation is slightly less rapid than the HEALPix implementation
with zero iteration ($i=0$), but becomes more rapid as soon as at
least one iteration ($i\geq1$) is considered in the HEALPix scheme.
The computation of the directional correlation of multiple signals
or simulations with steerable filters is thereby rendered easily affordable
at high resolutions.

In the perspective of the scale-space analysis of the CMB temperature
($T$) and polarization ($E$ and \textbf{$B$}) anisotropies, the
wavelet processing of the CMB data is therefore easily affordable
at the high resolutions of the WMAP and Planck experiments. The identification
of possible local non-Gaussianity or statistical anisotropy signatures,
or foreground emissions, is accessible with the same high precision
in both position and local direction on the sphere. The low computation
times and very good precision of the equi-angular and HEALPix implementations
of our algorithm was clearly illustrated in that context through a
wavelet decomposition of a simulated three-year WMAP temperature map
of several megapixels.

The generic algorithm developed for the fast directional correlation
on the sphere may also find other applications, notably in the analysis
of asymmetric beam effects on the CMB temperature and polarization
data, or well beyond cosmology.

\acknowledgements{The authors acknowledge the use of the LAMBDA archive, and of the
HEALPix and CMBFAST softwares. They wish wish to thank J.-P. Antoine,
B. Barreiro, and E. Mart\'inez-Gonz\'alez for valuable comments
and discussions. They acknowledge support of the HASSIP (Harmonic
Analysis and Statistics for Signal and Image Processing) European
research network (HPRN-CT-2002-00285). Y. W. acknowledges support
of the Swiss National Science Foundation (SNF) under contract No.
200021-107478/1. He is also postdoctoral researcher of the Belgian
National Science Foundation (FNRS). P. V. was supported by the Spanish
MEC project ESP2004-07067-C03-01.}

\end{document}